\documentclass[a4paper,floatfix,aps,pra,showpacs,twocolumn]{revtex4} 
\usepackage{amsmath}
\usepackage{amsfonts}
\usepackage{bm}
\usepackage{dcolumn}
\usepackage{graphicx}
\usepackage[english]{babel}

\begin{document}

\author{Beno{\^\i}t Gr{\'e}maud}
\email{Benoit.Gremaud@spectro.jussieu.fr}
\affiliation{Laboratoire Kastler Brossel, Universit{\'e} Pierre et
Marie Curie, T12, E1 \\
4, place Jussieu, 75252 Paris Cedex 05, France}
\title{$\hbar$ corrections in semiclassical formulas for smooth
  chaotic dynamics} 
\date{\today}

\begin{abstract}
The validity of semiclassical expansions in the power of $\hbar$ for the
quantum Green's function have been extensively 
tested for billiards systems, but in the case of chaotic
dynamics with smooth potential, even if formula are existing, a
quantitative comparison is still missing. In this paper, extending the 
theory developed by Gaspard
et~\textit{al.}, Adv.~Chem.~Phys. \textbf{XC} 105 (1995), based on 
the classical Green's functions, we present an efficient method 
allowing the calculation of 
$\hbar$ corrections for the propagator, the quantum Green's function,
and their traces. Especially, we show that the previously published
expressions for $\hbar$ corrections to the traces are incomplete.
\end{abstract} 

\maketitle

\section{Introduction}
Gutzwiller's work has now become a milestone in the understanding of the
properties of a quantum system whose classical counterpart depicts
chaotic dynamics~\cite{Gutzwiller90}. Starting from Feynman's path
formulation of quantum mechanics, he has been able to
complete the early studies of Van~Vleck~\cite{Vleck28},  
deriving expressions for the semiclassical propagator, and from this,
for the quantum level density: the well-known Gutzwiller trace
formula. The later is an asymptotic series in $\hbar$ and
can be separated into two parts; the leading order corresponds to the
Thomas-Fermi (or extended Thomas-Fermi when including $\hbar$
corrections) average density of states~\cite{BB97}; the other
part corresponds to the oscillations around the preceding term and
involves contributions from all periodic orbits of the system. This
formula has been widely used to obtain approximate values for the
quantum energy eigenvalues of classically chaotic systems: the
hydrogen atom in magnetic field~\cite{FW89,Houches89}, the helium
atom~\cite{Ezra91,GR93,GG98}, anisotropic Kepler 
problem~\cite{Gutzwiller90}, resonant tunnel diode~\cite{SM98},
billiards~\cite{Bu79,CE89,GA93,GAB95}, etc. Since then, the Gutzwiller 
trace formula has also been generalized to take into account
contributions of other kinds: diffractive effects~\cite{BGS99}, continuous
families of periodic orbits\cite{CL91,CL92,GAB95}, ghost orbits, etc. 

At the same time, because the trace formula as derived by Gutzwiller only
contained the leading term of the asymptotic expansion of the quantum level
density, the systematic expansion of the semiclassical propagator in
powers of $\hbar$ has been the purpose of several
studies~\cite{GA93,GAB95,VR96}. 
However, these corrections to the trace formula have only been tested
for billiards, for which both classical and quantum properties are
easier to calculate. In the present paper, we will show how, for
quantum systems whose Hamiltonian separates into kinetic and smooth
potential energies, $\hbar$ corrections can be computed with great
accuracy, extending the method described in Refs.~\cite{GA93,GAB95},
based on classical Green's functions. Especially, we will show that
the previous derivation~\cite{GA93,GAB95} of the correction to
Gutzwiller trace formula is partially wrong.

From a numerical point of view, 
all quantities involved in the calculation of the $\hbar$ corrections
for a given classical path can be obtained as solutions of sets of first order
differential equations to be integrated along this path using standard
time integrators like the Runge-Kutta method. The number of equations
in these sets can be quite large and can be probably reduced with a
deeper analysis of their structures, in the same way that the
amplitude in the Gutzwiller trace formula for a two-dimensional (2D)
system can be 
obtained by integrating only a $(2\times 2)$ matrix and not the whole
monodromy matrix~\cite{EW91}. However, it would give rise to more
complicated expressions and probably to additional difficulties in the
numerical implementation, whereas the expressions given in the paper can
be put in the computer as they stand. Also, the amount of CPU~time and
the memory needed by the codes are small enough, so that, on a first stage,
 the reduction of the number of equations can be skipped.

The paper is divided as follows: in Sec.~\ref{seckqq0}, after a brief
description of the derivation of $\hbar$ corrections for the propagator
$K(\mathbf{q},\mathbf{q}_0,T)$, we explain how the classical Green's
functions and $\hbar$ corrections can be efficiently
computed. In Sec.~\ref{secgqq0}, we explain how to obtain
expression of the additional terms, arising from the time to energy
domain transformation, in $\hbar$ corrections for
the quantum Green's function $G(\mathbf{q},\mathbf{q}_0,T)$. 
In the case of the trace of the propagator, a detailed derivation of
the $\hbar$ corrections is carried out in Sec.~\ref{sectrk},
leading to the proper formulas, along with the way they can be
computed. The time to energy transformation is explained in
Sec.~\ref{Sec:TG}, leading to the $\hbar$ correction expression in
the case of the quantum Green's function.
Finally, section~\ref{bidihy} shows how to apply 
theoretical expressions obtained in the four preceding sections in
the case of the 2D hydrogen in magnetic field and emphasizes the
excellent agreement with numerical coefficients extracted from exact
quantum calculation, using harmonic inversion~\cite{WMW98,M99,WMW00}.

\section{The Propagator $K(\mathbf{\lowercase{q}},\mathbf{\lowercase{q}}_0,T)$}
\label{seckqq0}

\subsection{Feynman path integral}

The starting point is the Feynman path integral, whose discrete
version, for a time independent Hamiltonian
which separates into kinetic and potential energies, 
$\hat{H}=\hat{\mathbf{p}}^2/{2}+V({\hat{{\mathbf{q}}}})$,
reads as follows~\cite{GAB95}: 
\begin{multline}
\label{FPI2}
K(\mathbf{q},\mathbf{q}_0,T)=\int
  d\mathbf{q}_1\,d\mathbf{q}_2\,,\cdots,d\mathbf{q}_{N-1}
  (2\pi i\hbar\Delta t)^{-Nf/2} \\
  \times\exp\left[\frac i{\hbar}\sum_{n=0}^{N-1}
    L\left(\frac{\mathbf{q}_{n+1}-\mathbf{q}_n}{\Delta
        t},\mathbf{q}_n\right)\Delta t+\mathcal{O}(\Delta t)\right],
\end{multline}
where $\Delta t=T/N$, $\mathbf{q}_N=\mathbf{q}$ and
$L(\dot{\mathbf{q}},\mathbf{q})$ is the classical Lagrangian.   

For small values of $\hbar$ (i.e., the semiclassical limit), using the
stationary phase approximation, all
preceding integrals are expanded around the stationary solutions, that
is the classical orbits $\mathbf{q}_l^{\mathrm{cl}}(t)$ going
from $\mathbf{q}_0$ to $\mathbf{q}$ during time $T$, each of them thus 
giving a contribution $K_l(\mathbf{q},\mathbf{q}_0,T)$ to the
propagator, whose final expression reads formally as follows~\cite{GAB95}:
\begin{equation}
K_l(\mathbf{q},\mathbf{q}_0,T)=K_l^{(0)}(\mathbf{q},\mathbf{q}_0,T)
\bigl\{1+i\hbar
C_1(\mathbf{q},\mathbf{q}_0,T)+\mathcal{O}(\hbar^2)\bigr\},
\end{equation}
where $K_l^{(0)}(\mathbf{q},\mathbf{q}_0,T)$ is the dominant
semiclassical contribution to the propagator 
$K(\mathbf{q},\mathbf{q}_0,T)$:
\begin{multline}     
\label{Kl0(T)}
  K_l^{(0)}(\mathbf{q},\mathbf{q}_0,T)=\\
\frac 1{(2\pi{i}\hbar)^{f/2}}
  \left|\det{\left(-\frac{\partial^2
}{\partial\mathbf{q}\partial\mathbf{q}_0}
W_l^{\mathrm{cl}}(\mathbf{q},\mathbf{q}_0,T)\right)}\right|^{1/2}\\
\times\exp{\left[\frac{i}{\hbar}W_l^{\mathrm{cl}}(\mathbf{q},\mathbf{q}_0,T)
-i\frac{\pi}2\nu_l\right]},
\end{multline}
where $W_l^{\mathrm{cl}}(\mathbf{q},\mathbf{q}_0,T)$ is the classical
action and $\nu_l$ is the Morse index of the orbit. The
$C_1(\mathbf{q},\mathbf{q}_0,T)$ expression is given by~\cite{GAB95}:
\begin{multline}
\label{C1(T)}
\frac{1}8\int_0^T\!\!dt\,V^{(4)}_{ijkl}(t)
\mathcal{G}_{ij}(t,t)\mathcal{G}_{kl}(t,t)\\
+\frac{1}{24}
\int_0^T\!\!\!\!\int_0^T\!\!dtdt'\,V^{(3)}_{ijk}(t)V^{(3)}_{lmn}(t')\\
\times\bigl[
3\mathcal{G}_{ij}(t,t)\mathcal{G}_{kl}(t,t')\mathcal{G}_{mn}(t',t')\bigr.\\
\bigl.+2\mathcal{G}_{il}(t,t')\mathcal{G}_{jm}(t,t')
\mathcal{G}_{kn}(t,t')\bigr],
\end{multline}
where the $V^{(n)}(t)$ are higher-order derivatives of the
potential $V$, evaluated at $\mathbf{q}_l^{\mathrm{cl}}(t)$. 

The classical Green's function $\mathcal{G}(t,t')$, associated with the 
classical orbit, is an $(f\times f)$ matrix solution of the following
equation~\cite{GAB95} :
\begin{equation}
\label{JHE}
\mathcal{D}\cdot\mathcal{G}(t,t')=\openone\,\delta(t-t'),
\end{equation}
where $\mathcal{D}$ is the Jacobi-Hill operator, controlling the
linear stability around the classical orbit in the 
configuration space~\cite{GAB95}:
\begin{equation}
\label{JHO}
\mathcal{D}=-\frac{\mathrm{d}^2}{\mathrm{d}t^2}\openone-
\frac{\partial^2V}{\partial\mathbf{q}\partial\mathbf{q}}
\left[\mathbf{q}^{\mathrm{cl}}(t)\right].
\end{equation}
Furthermore, the fact that both initial and final point are fixed in
the propagator $K(\mathbf{q},\mathbf{q}_0,T)$
imposes the following boundary conditions on the classical
Green's function~\cite{GAB95}: 
\begin{equation} 
\label{BCK}
\mathcal{G}(0,t')=\mathcal{G}(T,t')=0\quad \forall\,t'\in[0,T].
\end{equation}

\subsection{Classical Green's function}
\label{CGFKQQ0}
If $\mathbf{q}_{l}(T)$ is a conjugate point of  $\mathbf{q}_0$, then
the determinant
$\det{(-\partial^2_{{\mathbf{q}\mathbf{q}}_0}W_l^{\mathrm{cl}})}$ in
formula~(\ref{Kl0(T)}) is formally infinite, but this happens
only for restricted values of $T$, so that, 
in this section, we will focus on the general
case, for which $\mathbf{q}_l(T)$ and $\mathbf{q}_0$ are not
conjugate points.

Apart from $t=t'$, $\mathcal{G}(t,t')$ obeying the homogeneous
Jacobi-Hill equation
$\mathcal{D}\cdot\mathcal{G}=0$, so that, introducing the notations
\begin{equation}
\left\{\begin{aligned}
\mathcal{G}_-(t,t') &= \mathcal{G}(t,t') 
\quad \text{for } 0\leq t\leq t',  \\
\mathcal{G}_+(t,t') &= \mathcal{G}(t,t') 
\quad \text{for } t'\leq t\leq T ,
\end{aligned}\right.
\end{equation}
one immediately obtains
\begin{equation}
\left(\begin{array}{c}
\mathcal{G}{\pm}(t,t')\\
\dot{\mathcal{G}}_{\pm}(t,t')
\end{array} \right)
= M(t)
\left(\begin{array}{c} 
A_{\pm}(t') \\
B_{\pm}(t') 
\end{array} \right),
\end{equation}
where $M(t)$ is the $(2f\times 2f)$ monodromy matrix, depicting the linear
stability around the classical orbit in the phase space.
$A_{\pm}$  and $B_{\pm}$ are four  $(f\times f)$ matrices, whose
values are determined from the boundary conditions at time
$t=t'$:
\begin{equation}
\label{BCPM}
\left\{\begin{aligned}
\mathcal{G}_+(t',t')-\mathcal{G}_-(t',t')&=0,\\
\frac{d\mathcal{G}_-}{dt}(t',t')-\frac{d\mathcal{G}_+}{dt}(t',t')&=
\openone
\end{aligned}\right.
\end{equation}
and at times $t=0$ and $t=T$:
\begin{equation}
\label{BCPM0T}
\left\{\begin{aligned}
\mathcal{G}_-(0,t')&=0, \\
\mathcal{G}_+(T,t')&=0. 
\end{aligned}\right.
\end{equation}

\begin{figure}
\includegraphics[width=7cm,angle=-90]{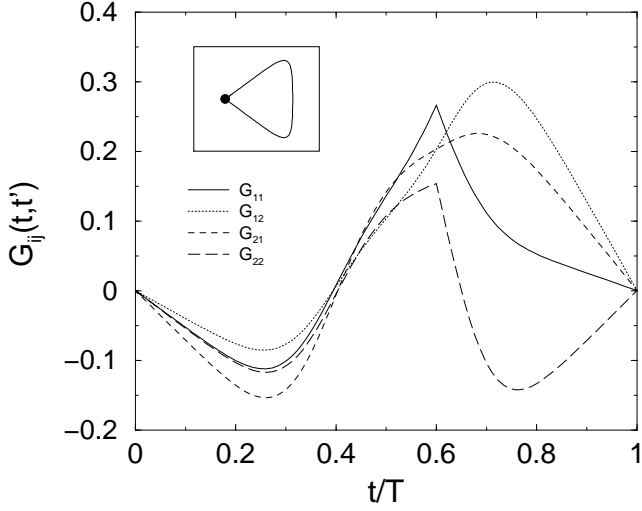}
\caption{\label{figg00} Example of a classical Green's function
  $\mathcal{G}(t,t')$ involved in the calculation of the $\hbar$
  corrections for the propagator $K(\mathbf{q},\mathbf{q}_0,T)$, for
  the case $\mathbf{q}=\mathbf{q}_0=\mathbf{0}$. It is
  associated 
  with the closed orbit $\overline{1243}$ of the 2D
  hydrogen atom in magnetic field, whose trajectory in $(u,v)$
  coordinates is inserted in the plot (see Sec.~\protect\ref{bidihy} for all
  details). This trajectory starts and ends at the nucleus, depicted
  by the black circle. Each curve corresponds to a matrix element
  $\mathcal{G}_{ij}(t,t')$ plotted with respect to time $t$, for
  $t'/T=0.6$. As expected from boundary
  conditions~(\protect\ref{BCK}), the Green's 
function vanishes at initial and  final times
(i.e. $\mathcal{G}(0,t')=\mathcal{G}(T,t')=0$) and for $t=t'$, the
derivatives of diagonal 
elements, $\mathcal{G}_{11}(t',t')$ (continuous line) and
$\mathcal{G}_{22}(t',t')$ (long dashed line), are
discontinuous whereas, from symmetry property~(\protect\ref{SPG})
(i.e., $\mathcal{G}^{\top}(t,t')=\mathcal{G}(t',t)$), the two off-diagonal
elements are equal (dotted and dashed lines).}
\end{figure}

For a Hamiltonian which separates between kinetic and potential energy
$H=\mathbf{p}^2/2+V(\mathbf{q})$, $M(t)$ has the following simple
structure:
\begin{equation}
M(t)=\left[\begin{array}{cc} 
J_2(t) & J_1(t) \\
\dot{J}_2(t) & \dot{J}_1(t)
\end{array} \right],
\end{equation}
which leads us  to the following explicit
expressions for the four matrices $A_{\pm}$ and $B_{\pm}$:
\begin{equation}
\left\{ \begin{array}{lcr}
A_-(t') & = & 0, \\
B_-(t') & = & J_2^{\top}(t')-J_1^{-1}(T)J_2(T)J_1^{\top}(t'),\\
A_+(t') & = & J_1^{\top}(t'), \\
B_+(t') & = & -J_1^{-1}(T)J_2(T)J_1^{\top}(t'),
\end{array}\right.
\end{equation}
provided that $J_1^{-1}(T)$ is invertible. $J_1(T)$ being the upper
right $(f\times f)$ submatrix of the matrix $M$, gives the linear
displacement of the final position for a change in the initial
momentum (the initial position being fixed to $\mathbf{q}_0$),
i.e., $\delta \mathbf{q}(T)=J_1(T)\delta \mathbf{p}_0$. Thus,
$J_1(T)$ is the inverse matrix of 
$(-\partial^2_{{\mathbf{q}\mathbf{q}}_0}W_l^{\mathrm{cl}})$
which has been supposed to be invertible 
($\mathbf{q}(T)$  and $\mathbf{q}_0$ are not conjugate
points). Finally the full expression for the
classical Green's function reads:
\begin{equation}
\label{GTTP}
\mathcal{G}(t,t')=\left\{\begin{array}{ccc}
J_1(t)&\left[J_2^{\top}(t')-J_1^{-1}(T)J_2(T)J_1^{\top}(t')\right]
& \\
& \mathrm{for} \quad 0\leq t\leq t', & \\
&&\\
&\left[J_2(t)-J_1(t)J_1^{-1}(T)J_2(T)\right] & J_1^{\top}(t') \\
&\mathrm{for} \quad t'\leq t\leq T. & \\
\end{array}\right.
\end{equation}
Using the symplectic structure of $M(T)$, one can show that
\begin{equation}
\label{SPG}
\mathcal{G}(t',t)=\mathcal{G}^{\top}(t,t')
\end{equation}
as expected because the operator $\mathcal{D}$ and the boundary
conditions are  symmetric as it
explicitly appears in the discrete version of the problem (see
Ref.~\cite{GAB95}). This is also emphasized in Fig.~\ref{figg00},
where the four matrix elements of a classical Green's function
$\mathcal{G}(t,t')$ (for $t'/T=0.6$) are plotted with respect to time
$t$. This example corresponds to a classical orbit of the 2D hydrogen
atom in a magnetic field having initial and final points on the nucleus,
namely, the closed orbit having  code $0-$ and whose trajectory in
$(u,v)$ coordinates is also shown in the figure. 
(see Sec.~\ref{bidihy} for all details.) As expected, the Green's
function vanishes at initial and  final times
(i.e., $\mathcal{G}(0,t')=\mathcal{G}(T,t')=0$) and for $t=t'$, the
derivatives of each diagonal 
elements $\mathcal{G}_{11}(t',t')$ (continuous line) and
$\mathcal{G}_{22}(t',t')$ (long dashed line) are
discontinuous whereas, from property~(\ref{SPG}), the two off-diagonal
elements are equal (dotted and dashed lines). 

\subsection{Getting $C_1(\mathbf{q},\mathbf{q}_0,T)$ by integrating
  a set of first order differential equations}
\label{C1qq0}

From Eq.~(\ref{C1(T)}), there are three contributions to 
$C_1(\mathbf{q},\mathbf{q}_0,T)$, namely:
\begin{widetext}
\begin{equation}
\begin{split}
I_1(T)&=\int_0^T\!\!dt\,\,V^{(4)}_{ijkl}(t)
\mathcal{G}_{ij}(t,t)\mathcal{G}_{kl}(t,t),\\
I_2^+(T)&=\int_0^T\!\!\!\!\int_0^T\!\!dtdt'\,\,V^{(3)}_{ijk}(t)
V^{(3)}_{lmn}(t')
\mathcal{G}_{ij}(t,t)\mathcal{G}_{kl}(t,t')\mathcal{G}_{mn}(t',t'),\\
I_2^-(T)&=\int_0^T\!\!\!\!\int_0^T\!\!dtdt'\,\,V^{(3)}_{ijk}(t)
V^{(3)}_{lmn}(t')
\mathcal{G}_{il}(t,t')\mathcal{G}_{jm}(t,t')\mathcal{G}_{kn}(t,t').
\end{split}
\end{equation}
\end{widetext}
Even if, in principle, one can compute $\mathcal{G}(t,t')$ for any
$(t,t')$ values using Eq.~(\ref{GTTP}), direct evaluation of the
double integrals $I_2^{\pm}$ would be time consuming and numerically
inefficient using standard integration routines, especially because,
from its definition, $\mathcal{G}(t,t')$ is not a 
smooth function around the line $t=t'$. In what follows, we will show
that the preceding integrals can be transformed in way such that their
values can be obtained integrating a set of first order differential
equations along the classical orbit, in the same way that, for example, the
monodromy matrix $M(T)$ can be computed.

Separating $t>t'$ and $t<t'$ contributions in $I_2^{\pm}$, using
symmetry property~(\ref{SPG}) of $\mathcal{G}(t,t')$ and that the 
matrix $V^{(3)}$ is fully symmetric under index permutations, one
gets, after straightforward algebra,
\begin{widetext}
\begin{equation}
\begin{split}
I_2^+(T)&=2\int_0^T\!\!dt\!\!\int_0^t\!\!dt'\,\,V^{(3)}_{ijk}(t)
V^{(3)}_{lmn}(t')
\mathcal{G}_{ij}(t,t)\mathcal{G}_{kl}(t,t')\mathcal{G}_{mn}(t',t'),\\
I_2^-(T)&=2\int_0^T\!\!dt\!\!\int_0^t\!\!dt'\,\,V^{(3)}_{ijk}(t)
V^{(3)}_{lmn}(t')
\mathcal{G}_{il}(t,t')\mathcal{G}_{jm}(t,t')\mathcal{G}_{kn}(t,t').
\end{split}
\end{equation}
\end{widetext}
In the preceding expressions the Green's function $\mathcal{G}(t,t')$
is used only for $(t,t')$ values in the triangle
$0\leq t'\leq t\leq T$ and is formally written 
$\mathcal{G}(t,t')=B_-^{\top}(t)J^{\top}_1(t')$ (see
Eq.~(\ref{GTTP})), thus separating $t$ and $t'$ contributions:
\begin{equation}
\begin{split}
I_2^+(T)&=2\int_0^Tdt\,\,V^{(3)}_{ijk}(t)\mathcal{G}_{ij}(t,t)B^-_{pk}(t)\\
&\phantom{=}\times
\int_0^tdt'\,\,V^{(3)}_{lmn}(t'){J_1}_{lp}(t')\mathcal{G}_{mn}(t',t'),\\
I_2^-(T)&=2\int_0^Tdt\,\,V^{(3)}_{ijk}(t)B^-_{pi}(t)B^-_{qj}(t)B^-_{rk}(t)\\
&\phantom{=}\times
\int_0^tdt'\,\,V^{(3)}_{lmn}(t'){J_1}_{lp}(t'){J_1}_{mq}(t'){J_1}_{nr}(t').
\end{split}
\end{equation}
This leads us to introduce two intermediate quantities, namely, $P_p(t)$
and $Q_{pqr}(t)$ (for $p$, $q$ and $r$ running from $1$ to $f$):
\begin{equation}
\begin{split}
P_p(t)&=\int_0^tdt'\,V^{(3)}_{lmn}(t'){J_1}_{lp}(t')\mathcal{G}_{mn}(t',t'),\\
Q_{pqr}(t)&=\int_0^tdt'\,V^{(3)}_{lmn}(t'){J_1}_{lp}(t'){J_1}_{mq}(t')
{J_1}_{nr}(t')
\end{split}
\end{equation}
in a way such that $I_2^{\pm}(T)$ (and $I_1(T)$) are solutions of the
following set of differential equations (besides equations for 
$\mathbf{X}(t)$ and $M(t)$):
\begin{equation}
\label{DSETKQQ0}
\left\{
\begin{aligned}
\dot{I}_1&=V^{(4)}_{ijkl}(t)\mathcal{G}_{ij}(t,t)\mathcal{G}_{kl}(t,t),\\
\dot{P}_p&=V^{(3)}_{lmn}(t){J_1}_{lp}(t)\mathcal{G}_{mn}(t,t), \\
\dot{I}_2^+&=V^{(3)}_{ijk}(t)\mathcal{G}_{ij}(t,t)B^-_{pk}(t)P_p(t,) \\
\dot{Q}_{pqr}&=V^{(3)}_{lmn}(t){J_1}_{lp}(t){J_1}_{mq}(t)
{J_1}_{nr}(t), \\
\dot{I}_2^-&=V^{(3)}_{ijk}(t)B^-_{pi}(t)B^-_{qj}(t)B^-_{rk}(t)Q_{pqr}(t) \\
\end{aligned}\right.
\end{equation}
with initial conditions $I_1(0)=I_2^{\pm}(0)=P_p(0)=Q_{pqr}(0)=0$.
This set of 
equations, $f^3+4f^2+3f+3$ in total (i.e., 33 for a 2D system) is
easily integrated using any standard method (fourth order
Runge-Kutta in the present case). As mentioned in the Introduction,
the size of the preceding differential set is probably
not minimal and could be reduced by a deeper analysis of the structure
of these equations. However, it allows a fast and easy computation of
the correction $C_1(\mathbf{q},\mathbf{q}_0,T)$: 
\begin{itemize}
\item[$\bullet$] Find a trajectory going from $\mathbf{q}_0$ to
  $\mathbf{q}$ in time $T$;
\item[$\bullet$] Integrate the differential
  set for $\mathbf{X}(t)$ and $M(t)$ along the trajectory to obtain
  the quantity   $J_1^{-1}(T)J_2(T)$;
\item[$\bullet$] Integrate the set of Eqs.~(\ref{DSETKQQ0})
  along the trajectory to get the three quantities $I_1$, $I_2^{\pm}$,
  entering in the $C_1(\mathbf{q},\mathbf{q}_0,T)$ expression.
\end{itemize}

\section{The Green's function
  $G(\mathbf{\lowercase{q}},\mathbf{\lowercase{q}}_0,E)$}  
\label{secgqq0}
\subsection{Going from time to energy domain}

since the quantum Green's function $G(\mathbf{q},\mathbf{q}_0,E)$ is
related to the propagator $K(\mathbf{q},\mathbf{q}_0,T)$, through
a semisided Fourier transform, this relation also holds between
semiclassical contributions arising from each classical orbit, more
precisely,
\begin{equation}
\label{GQQ0}
G_l(\mathbf{q},\mathbf{q}_0,E)=\frac 1{i\hbar}
\int_0^{+\infty} dT\,\exp{\left(\frac i{\hbar} ET\right)}
K_l(\mathbf{q},\mathbf{q}_0,T).
\end{equation}
Again, a stationary phase approximation is used to perform the integral,
which, for a given trajectory  going from $\mathbf{q}_0$ to
$\mathbf{q}$, selects its total duration $T_0$ such that the
classical motion is made at energy $E$. This operation also gives rise
to additional terms in $\hbar$ corrections, to be summed with
$C_1(\mathbf{q},\mathbf{q}_0,T)$, and whose explicit expressions
can be derived starting from Eq.~(\ref{C1(T)}) formally written as
follows~\cite{GAB95}: 
\begin{multline}
K_l(\mathbf{q},\mathbf{q}_0,T)=\frac{1}{(2\pi i\hbar)^{f/2}}
\exp\Bigl[\frac{i}{\hbar}
W_l(\mathbf{q},\mathbf{q}_0,T)-i\frac{\pi}2\nu_l\Bigr.\\
\Bigl.+C_0(\mathbf{q},\mathbf{q}_0,T)+i\hbar
C_1(\mathbf{q},\mathbf{q}_0,T)\Bigr],
\end{multline}
$C_0(\mathbf{q},\mathbf{q}_0,T)$ being thus the (logarithm of)
usual semiclassical amplitude. Then $W_l(\mathbf{q},\mathbf{q}_0,T)$ and 
$C_0(\mathbf{q},\mathbf{q}_0,T)$ are systematically expanded around
$T_0$:
\begin{equation}
\left\{
\begin{aligned}
W_l(\mathbf{q},\mathbf{q}_0,T)&=W_l^{(0)}+
\delta T\,W_l^{(1)}+\frac{\delta T^2}2W_l^{(2)}\\
&\phantom{=W_l^{(0)}}+\frac{\delta T^3}6
 W_l^{(3)}+\frac{\delta T^4}{24}W_l^{(4)}, \\
C_0(\mathbf{q},\mathbf{q}_0,T)&=C_0^{(0)}+
\delta T\,C_0^{(1)}+\frac{\delta T^2}2C_0^{(2)}
\end{aligned}\right.
\end{equation}
with $\delta T=(T-T_0)$. Terms arising from $C_1(T)$ expansion would
contribute only to $\hbar^2$ correction and can be
discarded. Performing the imaginary Gaussian integrals leads to the
additional $\hbar$ corrections:
\begin{multline}
C_1^{T\rightarrow E}(\mathbf{q},\mathbf{q}_0,T_0)=\frac{1}{2W_l^{(2)}}
\left[\left(C_0^{(1)}\right)^2+C_0^{(2)}\right]\\
-\frac{W_l^{(3)}C_0^{(1)}}{2\left(W_l^{(2)}\right)^2}
-\frac{W_l^{(4)}}{8\left(W_l^{(2)}\right)^2}+
\frac{5}{24}\frac{\left(W_l^{(3)}\right)^2}{\left(W_l^{(2)}\right)^3}.
\end{multline}
The preceding formula is similar to the one in Ref.~\cite{GAB95},
where the authors have expressed the coefficient 
$C_1^{T\rightarrow  E}(\mathbf{q},\mathbf{q}_0,T_0)$ in terms of
derivatives of amplitude and action with respect to energy $E$.  
The full expression of $G_l(\mathbf{q},\mathbf{q}_0,E)$ is then
given by
\begin{multline}
\label{gqqofin}
G_l(\mathbf{q},\mathbf{q}_0,E)=
\frac{2\pi}{\left(2\pi i\hbar\right)^{(f+1)/2}}
\frac{1}{\left|W_l^{(2)}\det{J_1(T_0)}\right|^{1/2}}\\
\times\exp{\left[\frac{i}{\hbar}S_l(\mathbf{q},\mathbf{q}_0,E)
-i\frac{\pi}2\tilde{\nu}_l\right]} \\
\times\Bigl\{1+i\hbar\left[C_1(\mathbf{q},\mathbf{q}_0,T_0)
+C_1^{T\rightarrow E}(\mathbf{q},\mathbf{q}_0,T_0)\right]
+\mathcal{O}(\hbar^2)\Bigr\},
\end{multline}
where $S_l(\mathbf{q},\mathbf{q}_0,E)$ is the reduced action and
\begin{equation}
\left\{\begin{array}{lr}
\tilde{\nu}_l=\nu_l &\mathrm{if}\quad W_l^{(2)}>0, \\
\tilde{\nu}_l=\nu_l +1&\mathrm{if}\quad W_l^{(2)}<0.
\end{array}\right.
\end{equation}

\subsection{Getting $C_1^{T\rightarrow E}(\mathbf{q},\mathbf{q}_0,T_0)$ by
  integrating a set of first order differential equations}
\label{C1TE}

In Sec.~\ref{C1qq0}, we have shown that 
$C_1(\mathbf{q},\mathbf{q}_0,T_0)$ can be computed
by integrating a set of differential equations along the classical
orbit going from $\mathbf{q}_0$ to $\mathbf{q}$ in time $T_0$. 
In this section we will show that it is also true for
$C_1^{T\rightarrow E}(\mathbf{q},\mathbf{q}_0,T_0)$, which
involves derivatives of both $W_l(\mathbf{q},\mathbf{q}_0,T)$ and 
$\det{J_1(T)}$ with respect to $T$.

For all $T$, we have the following functional relation 
($\mathbf{q}_0$ and $\mathbf{q}$ being fixed):
\begin{equation}
\label{dwldt}
\frac{\partial W_l(\mathbf{q},\mathbf{q}_0,T)}{\partial T}=
-E(\mathbf{q},\mathbf{q}_0,T),
\end{equation}
where $E(\mathbf{q},\mathbf{q}_0,T)$ is the energy of
the classical trajectory, $\mathbf{q}(t,T)$,
going from $\mathbf{q}_0$ to $\mathbf{q}$ in time
$T$, that is, the value of the Hamiltonian $H$ taken at any point on
the corresponding phase space trajectory 
$\mathbf{X}(t,T)=\bigl(\mathbf{q}(t,T),\mathbf{p}(t,T)\bigr)$. 

Writing $T=T_0+\delta T$, the Taylor expansion of
$H\bigl(\mathbf{X}(t,T_0+\delta T)\bigr)$ is easily deduced from 
the Taylor expansion of
$\mathbf{X}(t,T_0+\delta T)$ around the reference trajectory
$\mathbf{X}(t,T_0)$ (noted hereafter as $\mathbf{X}^{(0)}(t))$):
\begin{multline}
\label{TdT}
\mathbf{X}(t,T_0+\delta T)=\mathbf{X}^{(0)}(t)+
\delta T\,\mathbf{X}^{(1)}(t)\\
+\frac{\delta T^2}2\mathbf{X}^{(2)}(t)
+\frac{\delta T^3}6\mathbf{X}^{(3)}(t)+\cdots
\end{multline}
and from which one obtains the higher derivatives of the classical
action $W_l^{(n)}$ at $T=T_0$:
\begin{equation}
\label{dwdt}
\left\{\begin{aligned}
W_l^{(1)}&=-H\bigr(\mathbf{X}^{(0)}(t)\bigl), \\
W_l^{(2)}&=-X^{(1)}_iH^{(1)}_i, \\
W_l^{(3)}&=-\left(X^{(2)}_iH^{(1)}_i+X^{(1)}_iX^{(1)}_jH^{(2)}_{ij}\right),\\
W_l^{(4)}&=-\left(X^{(3)}_iH^{(1)}_i+3X^{(1)}_iX^{(2)}_jH^{(2)}_{ij}\right.\\
&\phantom{=-\left(X^{(3)}_iH^{(1)}_i\right.}
\left.+X^{(1)}_iX^{(1)}_jX^{(1)}_kH^{(3)}_{ijk}\right),
\end{aligned}\right.
\end{equation}
where all derivatives of $H$ are evaluated at $\mathbf{X}^{(0)}(t)$.

Equations for $\mathbf{X}^{(n)}(t)$ are deduced
from Hamilton's equations governing $\mathbf{X}(t,T)$ evolution:
\begin{equation}
\left\{\begin{aligned}
\label{dXdt}
\dot{X}_i^{(1)}&=\Sigma_{ij}H^{(2)}_{jk}X_k^{(1)},\\
\dot{X}_i^{(2)}&=\Sigma_{ij}H^{(2)}_{jk}X_k^{(2)}+
\Sigma_{ij}H^{(3)}_{jkl}X_k^{(1)}X_l^{(1)}, \\
\dot{X}_i^{(3)}&=\Sigma_{ij}H^{(2)}_{jk}X_k^{(3)}
+3\Sigma_{ij}H^{(3)}_{jkl}X_k^{(1)}X_l^{(2)}\\
&\phantom{=\Sigma_{ij}H^{(2)}_{jk}X_k^{(3)}}
+\Sigma_{ij}H^{(4)}_{jklm}X_k^{(1)}X_l^{(1)}X_m^{(1)},
\end{aligned}\right.
\end{equation}
where again all derivatives of $H$ are evaluated at
$\mathbf{X}^{(0)}(t)$. Thus, we are facing three differential sets
of the form  
$\dot{\mathbf{X}}^{(i)}=\Sigma
H^{(2)}\mathbf{X}^{(i)}+\Sigma\mathbf{Y}^{(i)}$ 
(i.e,. nonhomogeneous linear differential equations), with the
important property that the vector $\mathbf{Y}^{(i)}$ only depends on
vectors $\mathbf{X}^{(j)}$ with $j<i$, so that they can be solved
one after the other. Solutions of these nonhomogeneous linear 
differential equations are expressed with the monodromy matrix $M^{(0)}$:
\begin{equation}
\label{deltaXt}
\left\{\begin{aligned}
\mathbf{X}^{(1)}(t)&=M^{(0)}(t)\mathbf{X}^{(1)}(0,) \\
\mathbf{X}^{(2)}(t)&=M^{(0)}(t)\mathbf{X}^{(2)}(0)+\mathbf{F}^{(2)}(t),\\
\mathbf{X}^{(3)}(t)&=M^{(0)}(t)\mathbf{X}^{(3)}(0)+\mathbf{F}^{(3)}(t).
\end{aligned}\right.
\end{equation}
Among the $3\times(2f)$ dimensional space of solutions given by
preceding expressions, the relevant one is selected by transposing on 
initial values $\mathbf{X}^{(i)}(0)$ (for $i=1,2,3$) the two boundary
conditions 
\begin{equation}
\mathbf{q}(0,T_0+\delta T)=\mathbf{q}_0\quad\text{and}\quad
\mathbf{q}(T_0+\delta T,T_0+\delta T)=\mathbf{q}.
\end{equation}

Introducing position $\mathbf{q}^{(i)}$ and momentum
$\mathbf{p}^{(i)}$ parts for vectors $\mathbf{X}^{(i)}$, the Taylor
expansion of the preceding equations leads to the following boundary
conditions:
\begin{widetext}
\begin{equation}
\label{bct}
\left\{\begin{aligned}
\mathbf{q}^{(1)}(0)&=0\\
\mathbf{q}^{(2)}(0)&=0\\
\mathbf{q}^{(3)}(0)&=0
\end{aligned}\right.\qquad\text{and}\qquad
\left\{\begin{aligned}
\mathbf{q}^{(1)}(T_0)&=-\dot{\mathbf{q}}^{(0)}(T_0), \\
\mathbf{q}^{(2)}(T_0)&=-\ddot{\mathbf{q}}^{(0)}(T_0)-
2\dot{\mathbf{q}}^{(1)}(T_0), \\
\mathbf{q}^{(3)}(T_0)&=-\dddot{\mathbf{q}}^{(0)}(T_0)
-3\ddot{\mathbf{q}}^{(1)}(T_0)-3\dot{\mathbf{q}}^{(2)}(T_0).
\end{aligned}\right.
\end{equation}
\end{widetext}
Thus, the initial values $\mathbf{p}^{(i)}(0)$ are implicitly
determined by the final values $\mathbf{q}^{(i)}(T_0)$, through 
the integral
expressions~(\ref{deltaXt}), which for $\mathbf{X}^{(1)}$ reads
\begin{equation}
\label{XTX0}
\left(\begin{array}{c}
\mathbf{q}^{(1)}(T_0) \\
\mathbf{p}^{(1)}(T_0) \end{array}\right)=
\left[\begin{array}{cc}
J_2(T_0) & J_1(T_0) \\
\dot{J}_2(T_0) & \dot{J}_1(T_0) \end{array}\right]
\left(\begin{array}{c}
\mathbf{0} \\
\mathbf{p}^{(1)}(0)\end{array}\right)
\end{equation}
showing thus that
$\mathbf{p}^{(1)}(0)=-J_1^{-1}(T_0)\dot{\mathbf{q}}^{(0)}(T_0)$.

Then $\mathbf{F}^{(2)}(T_0)$ and $\mathbf{F}^{(3)}(T_0)$ are easily
computed by integrating sets of differential equations obtained from
Eq.~(\ref{dXdt}),  
allowing us to derive $\mathbf{p}^{(2)}(0)$ and $\mathbf{p}^{(3)}(0)$
values from 
Eq.~(\ref{deltaXt}), solving systems similar to Eq.~(\ref{XTX0}):
\begin{equation}
\left\{\begin{aligned}
\mathbf{p}^{(2)}(0)&=-J_1^{-1}(T_0)\left(\ddot{\mathbf{q}}^{(0)}(T_0)+
2\dot{\mathbf{q}}^{(1)}(T_0)+\mathbf{f}^{(2)}(T_0)\right),\\
\mathbf{p}^{(3)}(0)&=-J_1^{-1}(T_0)\left(\dddot{\mathbf{q}}^{(0)}(T_0)
+3\ddot{\mathbf{q}}^{(1)}(T_0)\right.\\
&\phantom{-J_1^{-1}(T_0)\left(\dddot{\mathbf{q}}^{(0)}(T_0)\right.}
\left.+3\dot{\mathbf{q}}^{(2)}(T_0)
+\mathbf{f}^{(3)}(T_0)\right),
\end{aligned}\right.
\end{equation}
where we have introduced the notation
$(\mathbf{f}^{(i)},\mathbf{g}^{(i)})$ for vectors
$\mathbf{F}^{(i)}$. Quantities like $\dot{\mathbf{q}}^{(1)}(T_0)$,
$\ddot{\mathbf{q}}^{(1)}(T_0)$, and $\dot{\mathbf{q}}^{(2)}(T_0)$ can
also be expressed in terms of $\mathbf{X}^{(0)}(T_0)$ and its derivatives.

At this point, from the values of the three vectors
$\mathbf{X}^{(i)}(T_0)$ and  using Eqs.~(\ref{dwdt}) at time $T_0$, 
all derivatives $W^{(n)}$ of the classical action can be computed.

We now explain how to
compute derivatives of $\det{J_1(T)}$. More precisely one has to
calculate the two coefficients $C_0^{(1)}$ and $C_0^{(2)}$, which are
derivatives of $-\ln{\sqrt{|\det{J_1(T)}|}}$, so that, using the
well-known formula
\begin{equation}
\label{dlndet}
\frac{d}{dT}\left(\ln{|\det{J}|}\right)=
\mathrm{Tr}\left(J^{-1}\frac{dJ}{dT}\right)
\end{equation}
($J$ being any (invertible) matrix), expressions of $C_0^{(1)}$ and $C_0^{(2)}$ become
\begin{equation}
\label{dc0dt}
\left\{\begin{aligned}
C_0^{(1)}&=-\frac{1}{2}\mathrm{tr}\left(J^{-1}_1(T_0)
\frac{dJ_1(T_0)}{dT}\right), \\
C_0^{(2)}&=-\frac{1}{2}\mathrm{tr}\left(J^{-1}_1(T_0)
\frac{d^2J_1(T_0)}{dT^2}\right.\\
&\phantom{=-\frac{1}{2}\mathrm{tr}\left(\right.}
\left.-J^{-1}_1(T_0)\frac{dJ_1(T_0)}{dT}
J^{-1}_1(T_0)\frac{dJ_1(T_0)}{dT}\right),
\end{aligned}\right.
\end{equation}
where $dJ_1(T_0)/dT$ means derivative of $J_1(T_0)$ when changing total
time $T$ (and thus the classical orbit), which must not be confused
with $\dot{J}_1$ (time derivative of $J_1$ along a given classical
orbit). $J_1(T)$ being the $(f\times f)$ upper right submatrix of the
monodromy matrix $M(T)$, $d^nJ_1(T_0)/dT^n$ is also stored at the
same position in matrix $d^nM(T_0)/dT^n$, for which we will derive
general expressions. For this purpose, we first introduce the explicit
notation $M(t,T)$, representing the value of the monodromy matrix at
time $t$ along the orbit going from $\mathbf{q}_0$ to $\mathbf{q}$
in time $T$. Writing $T=T_0+\delta T$, the Taylor expansion of
$M(t,T)$ for a given time $t$ reads
\begin{equation}
M(t,T_0+\delta T)=M^{(0)}(t)+\delta T\,M^{(1)}(t)
+\frac{\delta T^2}{2}M^{(2)}(t),
\end{equation}
where $M^{(0)}(t)$ is the monodromy matrix along the reference orbit
(i.e., going from $\mathbf{q}_0$ to $\mathbf{q}$ in time $T_0$).
Then $dM(T_0)/dT$ and $d^2M(T_0)/dT^2$ are the Taylor coefficients of
monodromy matrix $M(T_0+\delta T,T_0+\delta T)$  and thus have the
following expression: 
\begin{equation}
\label{dmdt}
\left\{\begin{aligned}
\frac{dM(T_0)}{dT}&=\dot{M}^{(0)}(T_0)+M^{(1)}(T_0), \\
\frac{d^2M(T_0)}{dT^2}&=\ddot{M}^{(0)}(T_0)+2\dot{M}^{(1)}(T_0)
+M^{(2)}(T_0).
\end{aligned}\right.
\end{equation}
Equations governing $M^{(i)}(t)$ evolution are easily deduced
from the one for $M(t,T)$:
\begin{equation}
\left\{\begin{aligned}
\dot{M}^{(1)}_{ij}&=\Sigma_{ik}\left[H^{(2)}_{kl}M^{(1)}_{lj}
+H^{(3)}_{klm} X^{(1)}_m M^{(0)}_{lj}\right],\\
\dot{M}^{(2)}_{ij}&=\Sigma_{ik}\left[H^{(2)}_{kl}M^{(2)}_{lj}
+2H^{(3)}_{klm} X^{(1)}_m M^{(1)}_{lj}\right.\\
&\phantom{=\Sigma_{ik}\left[\right.}
\left.+H^{(3)}_{klm} X^{(2)}_m M^{(0)}_{lj}
+H^{(4)}_{klmn} X^{(1)}_m X^{(1)}_n M^{(0)}_{lj}\right]
\end{aligned}\right.
\end{equation}
with initial conditions $M^{(1)}(0)=M^{(2)}(0)=0$. Obviously these
equations are similar to those governing $\mathbf{X}^{(i)}$
evolution, so that $M^{(1)}(T_0)$ and $M^{(2)}(T_0)$ values 
will be obtained by integrating similar differential sets. Actually,
it can be shown that
all these sets (for both $\mathbf{X}^{(i)}$ and $M^{(i)}$) can
be concatened in only one (larger) set of differential equations,
whose integration can be done at once. 

Finally, gathering all quantities in Eq.~(\ref{dmdt}), the two
matrices $dJ_1(T_0)/dT$ and $d^2J_1(T_0)/dT^2$ are inserted in
Eq.~(\ref{dc0dt}) thus giving values for 
$C_0^{(1)}$ and $C_0^{(2)}$, which, along with the values for
$W_l^{(n)}$, allow us to compute the numerical value for
$C_1^{T\rightarrow E}(\mathbf{q},\mathbf{q}_0,T_0)$. 

Obviously, the number of
equations in the preceding differential sets can be reduced, especially for
Hamiltonian separating into kinetic and potential energy, for which 
$H_{jkl}^{(3)}$ and $H_{jklm}^{(4)}$ coefficients are nonvanishing
only when $1\le j,k,l,m\le f$. However, these sets are straightforward to
implement and need only a small amount of CPU~time to be solved
using any conventional integrator ($4{\mathrm{th}}$ order Runge-Kutta in
the present case).

\section{Trace of the propagator $K(T)$}
\label{sectrk}

The diagonal elements $K(\mathbf{q}_0,\mathbf{q}_0,T)$ of the
propagator are related to classical orbits starting from
$\mathbf{q}_0$ and returning to this point after time $T$,
i.e., closed orbits. Summing
all these diagonal elements, that is performing the integral 
$\int d\mathbf{q}_0\,\,K(\mathbf{q}_0,\mathbf{q}_0,T)$, will
select, through another stationary phase approximation, closed orbits
for which initial and final momentum are equal: periodic
orbits. $\hbar$ corrections to leading order of the semiclassical
contribution to $K(T)$ from each periodic orbit can be derived
following the same scheme, previously used for the propagator
itself~\cite{Gutzwiller90,GAB95}.

\subsection{Feynman path integral}

Adding the integral over the initial and final positions in
Eq.~(\ref{FPI2}) yields~\cite{GAB95}
\begin{equation}
\label{TFPI}
\begin{split}
  K(T)&=\int d\mathbf{q}_0\,\,
  d\mathbf{q}_1\,\,d\mathbf{q}_2\,,\cdots,d\mathbf{q}_{N-1}
(2\pi i\hbar\Delta t)^{-Nf/2} \\
  &\phantom{=}\times\exp\left[\frac i{\hbar}\sum_{n=0}^{N-1}
    L\left(\frac{\mathbf{q}_{n+1}-\mathbf{q}_n}{\Delta
        t},\mathbf{q}_n\right)\Delta t+\mathcal{O}(\Delta t)\right]
\end{split}
\end{equation}
with $\mathbf{q}_N=\mathbf{q}_0$. 

The stationary phase approximation around a given periodic orbit
$\mathbf{q}_l^{\mathrm{cl}}(t)$ is made explicit when replacing the 
preceding $Nf$ integral with~\cite{GAB95}
\begin{equation}
\label{COORD}   
\int \mathrm{d}q_0^{\parallel}\,\,\mathrm{d}\bm{\xi}_0^{\perp}\,\,
  \mathrm{d}\bm{\xi}_1\,\,\mathrm{d}\bm{\xi}_2\,,\cdots,
 \mathrm{d}\bm{\xi}_{N-1}
\end{equation}
with
$\bm{\xi}_n=\mathbf{q}_n-\mathbf{q}_l^{\mathrm{cl}}(n\Delta t)$. 
For $n=0$ (i.e., initial position), only deviations perpendicular to the
periodic orbit $\bm{\xi}_0^{\perp}$ have been introduced
because the classical action  
$W_l(\mathbf{q}_0,\mathbf{q}_0,T)$ is constant along the orbit
(depicted by $q_0^{\parallel}$). The contribution $K_l(T)$ of this
periodic orbit to $K(T)$ then reads~\cite{GAB95}
\begin{equation}
\label{ETFPI}
\begin{aligned}
  K_l(T)=&
  \left(\frac{N}{2\pi{i}\hbar{T}}\right)^{Nf/2}\exp{\left(\frac{i}{\hbar}
    W_l\right)}
  \int \mathrm{d}q_0^{\parallel}\,\,\mathrm{d}\bm{\xi}_0^{\perp}\,\,
  \mathrm{d}\bm{\xi}_1\\
&\times\mathrm{d}\bm{\xi}_2\,,\cdots,
 \mathrm{d}\bm{\xi}_{N-1}\, 
  \exp{\left(\frac{i}{2\hbar}W_{,ab}\xi_a\xi_b\right)} \\
&\times\Bigl[1+\frac{i}{6\hbar}W_{,abc}\xi_a\xi_b\xi_c+
    \frac{i}{24\hbar}W_{,abcd}\xi_a\xi_b\xi_c\xi_d\Bigr.\\
&\Bigl.-\frac{1}{72\hbar^2}W_{,abc}W_{,def}\xi_a\xi_b\xi_c\xi_d\xi_e\xi_f
    \Bigr],
\end{aligned}
\end{equation}
where $\xi_a=\xi^{\perp}_{0i}$ when $a=(0,i)$ and $\xi_a=0$ when
$a=(0,0)$. $W_l$ is, in the large $N$ limit, the classical action of
the periodic orbit. Full expressions for $W_{,ab}$, $W_{,abc}$, and
$W_{,abcd}$ can be found in Ref.~\cite{GAB95}. 

Then, the next step would consist of performing all
imaginary Gaussian integrals, leaving out the integral along the
orbit. However, in the preceding coordinate
transformation~(\ref{COORD}),  there is 
an hidden subtlety, affecting only $\hbar$ corrections, which explains
why probably 
it is not mentioned in usual textbooks~\cite{Gutzwiller90,BB97}, where
authors are only looking at leading semiclassical amplitudes.

Actually, the problem is that the integral over $q_0^{\parallel}$
corresponds to the length of the classical orbit, only when  
$\bm{\xi}_0^{\perp}=\mathbf{0}$; for a nonzero value, it
will correspond to integration on a closed curve, slightly
displaced from the original trajectory, whose length will thus depend
on the $\bm{\xi}_0^{\perp}$ value. To enlighten this, let us
suppose that we have a bidimensional system, for which one periodic
orbit is a circle of radius $R_0$, traveled at constant speed 
$V_0=2\pi R_0/T$.  
The coordinate transformation is then easily made using polar coordinates
$(r,\theta)$:
\begin{equation}
r=R_0-\xi_0^{\perp}.
\end{equation}
The negative sign appears to preserve orientation. The volume element
$dx\,\,dy$ becomes
\begin{equation}
dx\,\,dy=rd\theta\,dr=(R_0-\xi_0^{\perp})d\theta\,d\xi_0^{\perp},
\end{equation}
which shows that, in this case, $dq_0^{\parallel}$ is not simply 
$R_0d\theta$, the length on the periodic orbit, but is given by
\begin{equation}
\label{dqcirc}
dq_0^{\parallel}=(R_0-\xi_0^{\perp})d\theta\neq R_0d\theta.
\end{equation}
This simple example shows actually that the variable $q_0^{\parallel}$
is not independent of $\bm{\xi}_0^{\perp}$, whereas
$\theta$ is. 

For a general system, the variable that can play the $\theta$ role is
actually the time $t$, whose variation domain $[0,T]$ is fixed and
then obviously independent of $\bm{\xi}_0^{\perp}$. Thus one
has to generalize the relation
$dq_0^{\parallel}=|\dot{\mathbf{q}}^{\mathrm{cl}}|dt_0$, valid only on the
periodic orbit. This is done by writing explicitly the coordinate
transformation
$\mathbf{q}\rightarrow(t_0,\bm{\xi}_0^{\perp})$:
\begin{equation}
\mathbf{q}=\mathbf{q}^{\mathrm{cl}}(t_0)+\xi_{0i}^{\perp}\,\mathbf{n}_i(t_0),
\end{equation}
where $\mathbf{n}_i(t_0)$ are $f-1$ orthogonal unit vectors lying
in the plane perpendicular to the periodic orbit at time $t_0$. The
Jacobian of the transformation reads
\begin{equation}
\begin{aligned}
\label{detcarloc}
\det{\frac{\partial\mathbf{q}}{\partial(t_0,\bm{\xi}_0^{\perp})}}&=\det{
\left[
\dot{\mathbf{q}}^{\mathrm{cl}}+\xi_{0i}^{\perp}\,\dot{\mathbf{n}}_i,  
\mathbf{n}_1,  \cdots, \mathbf{n}_{f-1}\right]}\\
&=|\dot{\mathbf{q}}^{\mathrm{cl}}|-\frac{1}{|\dot{\mathbf{q}}^{\mathrm{cl}}|}
\bm{\xi}_0^{\perp}\cdot\ddot{\mathbf{q}}^{\mathrm{cl}}.
\end{aligned}
\end{equation}

Inserting the volume element in Eq.~(\ref{ETFPI}), the contribution
$K_l(T)$ of the periodic orbit now reads, keeping only terms giving
rise to $\hbar$ corrections,
\begin{widetext}
\begin{multline}
\label{ETFPI3}
K_l(T)=
  \left(\frac{N}{2\pi{i}\hbar{T}}\right)^{Nf/2}\exp{\left(\frac{i}{\hbar}
    W_l\right)} 
  \int |\dot{\mathbf{q}}^{\mathrm{cl}}| dt_0\,\,
  \mathrm{d}\bm{\xi}_0^{\perp}\,\,
  \mathrm{d}\bm{\xi}_1\,\,\mathrm{d}\bm{\xi}_2\,,\cdots,
 \mathrm{d}\bm{\xi}_{N-1} 
  \exp{\left(\frac{i}{2\hbar}W_{,ab}\xi_a\xi_b\right)}\\
  \left[1+\frac{\xi_{\tilde{d}}V_{,\tilde{d}}}
    {|\dot{\mathbf{q}}^{\mathrm{cl}}|^2}
    +\frac{i}{6\hbar}W_{,abc}\xi_a\xi_b\xi_c+
    \frac{i}{24\hbar}W_{,abcd}\xi_a\xi_b\xi_c\xi_d
    +\frac{i}{6\hbar}\frac{V_{,\tilde{d}}W_{,abc}
      \xi_{\tilde{d}}\xi_a\xi_b\xi_c}
    {|\dot{\mathbf{q}}^{\mathrm{cl}}|^2}
       -\frac{1}{72\hbar^2}W_{,abc}W_{,def}\xi_a\xi_b\xi_c\xi_d\xi_e\xi_f
    \right],
\end{multline}
\end{widetext}
where we have seen that 
$\ddot{\mathbf{q}}^{\mathrm{cl}}=-\partial_{\mathbf{q}}V$ and we have
introduced the index $\tilde{d}$ for $(0,j)$.

As explained in Ref.~\cite{GAB95}, the imaginary Gaussian integrals
can be expressed in terms 
of another classical Green's functions $\mathcal{G}(t,t')$,
whose boundary conditions are extracted when comparing the detailed
expression of $W_{,ab}$ with the discrete version of the Jacobi-Hill
operator $\mathcal{D}$, see Eq.~(\ref{JHO}). Especially, it can be
shown that, in the large $N$ limit, they become 
\begin{equation}
\label{BCP}
\left\{\begin{aligned}
\mathcal{G}(0,t')&=\mathcal{G}(T,t'), \\
\mathcal{P}_{t_0}\mathcal{G}(0,t') &=
\mathcal{P}_{t_0}\mathcal{G}(T,t') =0, \\
\mathcal{Q}_{t_0}\dot{\mathcal{G}}(0,t') &=
\mathcal{Q}_{t_0}\dot{\mathcal{G}}(T,t'),
\end{aligned}\right.\quad \forall\,t'\in[0,T],
\end{equation}
where we have introduced $\mathcal{P}_{t_0}$ the
projector along the periodic orbit at time $t_0$ and
$\mathcal{Q}_{t_0}=\openone-\mathcal{P}_{t_0}$. In Ref.~\cite{GAB95},
only the $f^2+f$ boundary conditions corresponding to the first two
lines were given, whereas the $f^2-f$ ones corresponding to the last line
were missing. 

Gathering all results and taking the large $N$ limit in
Eq.~(\ref{ETFPI3}), the contribution of the given periodic orbit to
the trace of the propagator reads as follows:
\begin{equation}
\label{TKT}
K_l(T)=K_l^{(0)}(T)
\left\{1+i\hbar \frac{1}{T}\int_0^Tdt_0\,C_1(T,t_0)
+\mathcal{O}(\hbar^2)\right\},
\end{equation}
$K_l^{(0)}(T)$ being the usual semiclassical leading
order~\cite{Gutzwiller90,GAB95,L31}
\begin{multline}
K_l^{(0)}(T)=\frac{1}{\sqrt{2\pi\hbar}}\frac{T}{\left|\partial_ET
\det{(m(T)-\openone)}\right|^{1/2}}\\
\times\exp{\left[\frac{i}{\hbar}W_l(T)-i\frac{\pi}2\mu_l
+i\,\mathrm{sgn}\,\,\partial_ET\right]},
\end{multline}
where $W_l(T)$ is the classical action of the periodic orbit and
$\mu_l$ its Maslov index. 

The first $\hbar$ correction $C_1(T)$ to 
$K_l^{(0)}(T)$ is then obtained by averaging over the time $t_0$
(i.e. over the full periodic orbit) the coefficient $C_1(T,t_0)$,
given by
\begin{widetext}
\begin{multline}
C_1(T,t_0)=\frac{1}{8}\int_0^T\!\!dt\,\,
V^{(4)}_{ijkl}(t)\mathcal{G}_{ij}(t,t)\mathcal{G}_{kl}(t,t)
+\frac{1}{2}\frac{V^{(1)}_l(t_0)}{|\dot{\mathbf{q}}^{\mathrm{cl}}|^2}
\int_0^T\!\!dt\,\,V^{(3)}_{ijk}(t)\mathcal{G}_{lk}(0,t)\mathcal{G}_{ij}(t,t)\\
+\frac{1}{24}
\int_0^T\!\!\!\!\int_0^T\!\!dtdt'\,\,V^{(3)}_{ijk}(t)V^{(3)}_{lmn}(t')
\bigl[
3\mathcal{G}_{ij}(t,t)\mathcal{G}_{kl}(t,t')\mathcal{G}_{mn}(t',t')
+2\mathcal{G}_{il}(t,t')\mathcal{G}_{jm}(t,t')
\mathcal{G}_{kn}(t,t')\bigr],
\end{multline}
\end{widetext}
where $t_0$ represents thus the position $\mathbf{q}_0$ on the
periodic orbit at 
which boundary conditions~(\ref{BCP}) on the classical Green's function
$\mathcal{G}(t,t')$ are applied. $\mathbf{q}_0$ is also the initial
(and final) position on the periodic orbit for classical motions
corresponding to times $t$ and $t'$ entered in the
preceding expression.

\subsection{Classical Green's function}

As in Sec.~\ref{CGFKQQ0}, where expressions for classical Green's
functions  for the propagator $K(\mathbf{q},\mathbf{q}_0,T)$ where
derived, we introduce the $\mathcal{G}_{\pm}(t,t')$ notations and 
$A_{\pm}(t')$,  $B_{\pm}(t')$ matrices. Using all boundary
conditions (at times $t=t'$, $t=0$ and $t=T$) 
gives rise to the following equation:
\begin{multline}
\label{LSAB}
\left[\begin{array}{cc}
\openone & 0 \\
0 & \mathcal{Q}_{t_0}
\end{array}\right]\left(M(T)-\openone_{2f}\right)
\left(\begin{array}{c}
A_-(t') \\
B_-(t') 
\end{array}\right)=\\
\left[\begin{array}{cc}
\openone & 0 \\
0 & \mathcal{Q}_{t_0}
\end{array}\right]M(T)
\left(\begin{array}{r}
-J^{\top}_1(t')\\
J^{\top}_2(t')
\end{array}\right).
\end{multline}
The preceding set of linear equations, formally written
$\mathcal{A}\,\mathcal{X}=\mathcal{B}$, cannot be solved directly
because the $(2f\times 2f)$ matrix $\mathcal{A}$ is obviously
singular. More precisely, existence and number of solutions for the system 
$\mathcal{A}\,\mathbf{x}=\mathbf{b}$ are determined by the two
following properties:
\begin{itemize}
\item Solutions exists if for all vectors 
$\mathbf{y}$ such that $\mathcal{A}^{\top}\mathbf{y}=\mathbf{0}$, then
$\mathbf{y}\cdot\mathbf{b}=0$ .

\item If the preceding condition is fulfilled, and if $\mathbf{x}$
  is a solution, then for all vectors $\mathbf{x}_0$ such that
  $\mathcal{A}\,\mathbf{x}_0=\mathbf{0}$, $\mathbf{x}+\mathbf{x}_0$
  is also a solution, showing that the dimension of the solution space
  is that of the nullspace of $\mathcal{A}$.
\end{itemize}

In the present case, equation
$\mathbf{A}^{\top}\,\mathbf{y}=\mathbf{0}$ leads to, either
\begin{equation}
\left[\begin{array}{cc}
\openone & 0 \\
0 & \mathcal{Q}_{t_0}
\end{array}\right]\mathbf{y}=\mathbf{0} \qquad \Rightarrow\qquad
\mathbf{y}\propto
\left(\begin{array}{c}
\mathbf{0} \\
\dot{\mathbf{q}}(t_0)
\end{array}\right)
\end{equation}
or 
\begin{equation}
\bigl(M(T)-\openone_{2f}\bigr)\tilde{\mathbf{y}}=\mathbf{0}
\text{ with } 
\tilde{\mathbf{y}}=\Sigma
\left[\begin{array}{cc}
\openone & 0 \\
0 & \mathcal{Q}_{t_0}
\end{array}\right]\mathbf{y}\neq\mathbf{0}.
\end{equation}
For a generic unstable periodic orbit, the eigenspace associated with the
eigenvalue $1$ of $M(T)$ (T being the period), is of dimension one and
is spanned by the vector parallel to the flow (see the 
Appendix)
$\dot{\mathbf{X}}(t_0)=\left(\dot{\mathbf{q}}(t_0),\dot{\mathbf{p}}(t_0)\right)$,
so that, in the second case, one gets
$\tilde{\mathbf{y}}\propto\dot{\mathbf{X}}(t_0)$ and $\mathbf{y}$ is
a solution of: 
\begin{equation}
\left[\begin{array}{cc}
\openone & 0 \\
0 & \mathcal{Q}_{t_0}
\end{array}\right]\mathbf{y}\propto
\left(\begin{array}{r}
-\dot{\mathbf{p}}(t_0) \\
\dot{\mathbf{q}}(t_0)
\end{array}\right),
\end{equation}
which is impossible unless $\dot{\mathbf{q}}(t_0)=\mathbf{0}$, which, for
Hamiltonian separating into kinetic and potential energies,
corresponds to a self-retracing periodic orbit, for which a slightly
modified approach should be developed~\cite{EW91}. Nevertheless, this case is
peculiar, and we will suppose in the rest of the section that
$\dot{\mathbf{q}}$ never vanishes along the
periodic orbit in consideration.

Thus, the nullspace of $\mathcal{A}^{\top}$ being one-dimensional and 
spanned by the vector $\left(\mathbf{0},\dot{\mathbf{q}}(t_0)\right)$,
Eq.~(\ref{LSAB}) immediately shows that for any column of matrix
$\mathcal{B}$, we get 
$\left(\mathbf{0},\dot{\mathbf{q}}(t_0)\right)\cdot\mathcal{B}_i=0$,
thus fulfilling
the first condition. Denoting $\mathcal{X}_0$ as a solution of
Eq.~(\ref{LSAB}), which can be easily obtained using singular value
decomposition (SVD)
of matrix $\mathcal{A}$, and the nullspace of $M(T)-\openone$ being
spanned by $\dot{\mathbf{X}}(t_0)$, the general solution of 
Eq.~(\ref{LSAB}) reads
\begin{equation}
\label{solgen}
\mathcal{X}=\mathcal{X}_0+
\left[\alpha_1\dot{\mathbf{X}}(t_0),\alpha_2\dot{\mathbf{X}}(t_0),\cdots,
\alpha_f\dot{\mathbf{X}}(t_0)\right],
\end{equation}
where $\alpha_i$ are  unknown real parameters still to be
determined. Actually, in Eq.~(\ref{LSAB}) one boundary condition has
not been taken into account, namely that 
$\mathcal{P}_{t_0}\mathcal{G}_-(0,t')=0$ which, using
that the projector $\mathcal{P}_{t_0}$ reads
\begin{equation}
(\mathcal{P}_{t_0})_{ij}=\left(\frac{\dot{\mathbf{q}}(t_0)
\dot{\mathbf{q}}^{\top}(t_0)}
{|\dot{\mathbf{q}}(t_0)|^2}\right)_{ij}=
\frac{\dot{\mathbf{q}}_i(t_0)
\dot{\mathbf{q}}_j(t_0)}
{|\dot{\mathbf{q}}(t_0)|^2}
\end{equation}
allows us to get $\alpha_i$ values and, from that, the final
expression
\begin{equation}
\label{Mproj}
\left(\begin{array}{c}
A_-(t') \\
B_-(t')
\end{array}\right)=\mathcal{X}_0-\frac{1}{|\dot{\mathbf{q}}(t_0)|^2}
\left[\begin{array}{cc}
\dot{\mathbf{q}}(t_0)\dot{\mathbf{q}}^{\top}(t_0) & 0 \\
\dot{\mathbf{p}}(t_0)\dot{\mathbf{q}}^{\top}(t_0) & 0
\end{array}\right]\mathcal{X}_0,
\end{equation}
which, of course, is now independent of the particular solution
$\mathcal{X}_0$.

\begin{figure}
\includegraphics[width=7cm,angle=-90]{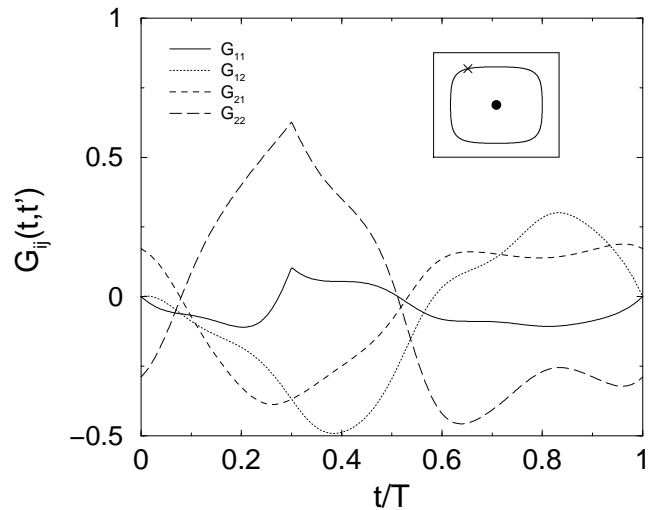}
\caption{\label{figtrg} Example of a classical Green's function
  $\mathcal{G}(t,t')$ involved in the calculation of the $\hbar$
  corrections for the trace of the propagator $K(T)$. It is associated
  with the periodic orbit $\overline{1234}$ of the 2D
  hydrogen atom in a magnetic field, whose trajectory in $(u,v)$
  coordinates is inserted in the plot (see Sec.~\protect\ref{bidihy} for all
  details). The black circle depicts the nucleus, whereas the cross
  corresponds to the initial and final points on the periodic orbit at
  which $\mathcal{G}(t,t')$ fulfills the boundary
  conditions~(\protect\ref{BCP}). Each curve corresponds to a matrix element
  $\mathcal{G}_{ij}(t,t')$ plotted with respect to time $t$, for $t'/T=0.3$.
  Actually, we have plotted the coefficient of the rotated matrix
$\tilde{\mathcal{G}}(t,t')$, such that its first row corresponds to the
direction parallel to the orbit; $\tilde{\mathcal{G}}_{11}(t,t')$
(continuous line) and $\tilde{\mathcal{G}}_{12}(t,t')$ (dotted line)
are thus equal to zero for initial ($t=0$) and final ($t=T$)
points. The other boundary conditions can also be verified in the
figure: the dashed line ($\tilde{\mathcal{G}}_{21}(t,t')$)
(respectively the long dashed line, $\tilde{\mathcal{G}}_{22}(t,t')$) has not
only the same value at initial and final time, but also the same
slope, which means that $\tilde{\mathcal{G}}_{21}(t,t')$ 
(respectively. $\tilde{\mathcal{G}}_{22}(t,t')$) and its time derivative
fulfills the periodic boundary conditions~(\protect\ref{BCP}). Finally,
for $t=t'$,
the off-diagonal coefficients $\tilde{\mathcal{G}}_{12}(t',t')$
(dotted line) and $\tilde{\mathcal{G}}_{21}(t',t')$ (dashed line) are
equal, as expected from the symmetry property
$\mathcal{G}^{\top}(t,t')=\mathcal{G}(t',t)$ .}
\end{figure}

Whereas in the case of the propagator $K(\mathbf{q},\mathbf{q}_0,T)$,
for which we were able to give an explicit expression~(\ref{GTTP}),
the classical 
Green's function associated with the trace of the propagator $K(T)$ is
only defined trough a linear system~(\ref{LSAB}), which nevertheless
allows us to obtain its numerical value for any $(t,t')$. Although it
clearly appears that matrix $\mathbb{D}$ expression~(see Ref.~\cite{GAB95}) is
symmetric, meaning that the classical Green's
function must fulfill the property
$\mathcal{G}^{\top}(t,t')=\mathcal{G}(t',t)$, getting the later directly from
Eq.~(\ref{LSAB}) is not obvious. However, in the case of the 2D
hydrogen in a magnetic field (see Sec.~\ref{bidihy} for all details),
we have numerically checked that the property holds. For example, in
Fig.~\ref{figtrg}  the four coefficients of
classical Green's function $\mathcal{G}(t,t')$ (for $t'/T=0.3$) of the
periodic orbit $1234$ are plotted with respect to time $t$. The
starting point $t_0$ on the periodic orbit is depicted by the cross.
Actually, we have plotted the coefficient of the rotated matrix
$\tilde{\mathcal{G}}(t,t')$, such that its first row corresponds to the
direction parallel to the orbit; $\tilde{\mathcal{G}}_{11}(t,t')$
(continuous line) and $\tilde{\mathcal{G}}_{12}(t,t')$ (dotted line)
are thus equal to zero for initial ($t=0$) and final ($t=T$)
points. The other boundary conditions can also be verified in the
figure: the dashed line ($\tilde{\mathcal{G}}_{21}(t,t')$)
(respectively the long dashed line, $\tilde{\mathcal{G}}_{22}(t,t')$) has not
only the same value at initial and final time, but also the same
slope, which means that $\tilde{\mathcal{G}}_{21}(t,t')$ 
(respectively $\tilde{\mathcal{G}}_{22}(t,t')$) and its time derivative
fulfill the periodic boundary conditions~(\ref{BCP}). Finally,
for $t=t'$,
the off-diagonal coefficients $\tilde{\mathcal{G}}_{12}(t',t')$
(dotted line) and $\tilde{\mathcal{G}}_{21}(t',t')$ (dashed line) are
equal, as expected from the symmetry property.

\subsection{Getting $C_1(T,t_0)$ by integrating a set of first order
  differential equations}
\label{C1Tt0}

As seen previously (see Sec.~\ref{C1qq0}), we will explain how the
numerical value of coefficients $C_1(T,t_0)$ can be obtained by
integrating a set of differential equation, using the standard Runge-Kutta
method. There are now four contributions to $C_1(T,t_0)$, namely

\begin{equation}
\begin{split}
I_1(T)=&\int_0^T\!\!dt\,\,V^{(4)}_{ijkl}(t)
\mathcal{G}_{ij}(t,t)\mathcal{G}_{kl}(t,t)\\
I_l(T)=&\int_0^T\!\!dt\,\,V^{(3)}_{ijk}(t)
\mathcal{G}_{lk}(0,t)\mathcal{G}_{ij}(t,t)\\
I_2^+(T)=&\int_0^T\!\!\!\!\int_0^T\!\!dt\,dt'\,\,V^{(3)}_{ijk}(t)V^{(3)}_{lmn}
(t')\\
&\times\mathcal{G}_{ij}(t,t)\mathcal{G}_{kl}(t,t')\mathcal{G}_{mn}(t',t')\\
I_2^-(T)=&\int_0^T\!\!\!\!\int_0^T\!\!dt\,dt'\,\,V^{(3)}_{ijk}(t)V^{(3)}_{lmn}
(t')\\
&\times\mathcal{G}_{il}(t,t')\mathcal{G}_{jm}(t,t')\mathcal{G}_{kn}(t,t').
\end{split}
\end{equation}

\begin{widetext}
The two main difficulties now are that $\mathcal{G}(t,t')$ does not
factorize anymore in a product of matrix at time $t$ and a matrix at
time $t'$, nor does the symmetric property
$\mathcal{G}^{\top}(t,t')=\mathcal{G}(t',t)$ explicitly appear (even
if we have numerically checked that it is fulfilled). Nevertheless,
as seen previously, separating $(t>t')$ and $(t<t')$ contributions in
$I_2^{\pm}(T)$ expressions and introducing four quantities $P^{(i)}_p$
for $1\le i\le4$, 
allows us to compute $I_2^+$, by integrating the following set of
differential equations from $t=0$ to $T$ (besides equations for 
$\mathbf{X}(t)$ and $M(t)$):
\begin{equation}
\left\{\begin{aligned}
\dot{P}_p^{(1)}&=A^+_{pl}(t)V^{(3)}_{lmn}(t)\mathcal{G}_{mn}(t,t),\qquad
\dot{P}_p^{(3)}={J_1}_{lp}(t)V^{(3)}_{lmn}(t)\mathcal{G}_{mn}(t,t),\\
\dot{P}_p^{(2)}&=B^+_{pl}(t)V^{(3)}_{lmn}(t)\mathcal{G}_{mn}(t,t),\qquad
\dot{P}_p^{(4)}={J_2}_{lp}(t)V^{(3)}_{lmn}(t)\mathcal{G}_{mn}(t,t),\\
\dot{I}_2^+&=V^{(3)}_{ijk}(t)\mathcal{G}_{ij}(t,t){J_2}_{kp}(t)P^{(1)}_p(t)+
V^{(3)}_{ijk}(t)\mathcal{G}_{ij}(t,t){J_1}_{kp}(t)P^{(2)}_p(t) \\
&+V^{(3)}_{ijk}(t)\mathcal{G}_{ij}(t,t)A^-_{pk}(t)P^{(3)}_p(t) 
+V^{(3)}_{ijk}(t)\mathcal{G}_{ij}(t,t)B^-_{pk}(t)P^{(4)}_p(t)
\end{aligned}\right.
\end{equation}
with vanishing initial conditions for $P^{(i)}_p$ and $I_2^+$. For
each time step, one must compute matrices $A_-$ and $B_-$ (and from there
matrices $A_+$ and $B_+$), solving the
linear system described in the previous section, using singular value
decomposition of matrix $\mathcal{A}$, which, being independent of
$t$, is done before starting the Runge-Kutta integration. 
Skipping intermediate steps, the differential equations leading to
$I_2^-(T)$ computation reads as follow, introducing another eight
quantities $Q^{(i)}_{pqr}$:
\begin{equation}
\label{DSETTRK}
\left\{\begin{aligned}
\dot{Q}^{(1)}_{pqr}&=V^{(3)}_{lmn}(t)A^+_{pl}(t)A^+_{qm}(t)A^+_{rn}(t),\qquad
\dot{Q}^{(5)}_{pqr}=V^{(3)}_{lmn}(t){J_2}_{lp}(t){J_2}_{mq}(t){J_2}_{nr}(t),\\
\dot{Q}^{(2)}_{pqr}&=V^{(3)}_{lmn}(t)A^+_{pl}(t)A^+_{qm}(t)B^+_{rn}(t),\qquad
\dot{Q}^{(6)}_{pqr}=V^{(3)}_{lmn}(t){J_2}_{lp}(t){J_2}_{mq}(t){J_1}_{nr}(t),\\
\dot{Q}^{(3)}_{pqr}&=V^{(3)}_{lmn}(t)A^+_{pl}(t)B^+_{qm}(t)B^+_{rn}(t),\qquad
\dot{Q}^{(7)}_{pqr}=V^{(3)}_{lmn}(t){J_2}_{lp}(t){J_1}_{mq}(t){J_1}_{nr}(t),\\
\dot{Q}^{(4)}_{pqr}&=V^{(3)}_{lmn}(t)B^+_{pl}(t)B^+_{qm}(t)B^+_{rn}(t),\qquad
\dot{Q}^{(8)}_{pqr}=V^{(3)}_{lmn}(t){J_1}_{lp}(t){J_1}_{mq}(t){J_1}_{nr}(t),\\
\dot{I}_2^-&=\phantom{3}V^{(3)}_{ijk}(t){J_2}_{ip}(t){J_2}_{jq}(t){J_2}_{kr}(t)
Q^{(1)}_{pqr}(t) 
+3V^{(3)}_{ijk}(t){J_2}_{ip}(t){J_2}_{jq}(t){J_1}_{kr}(t)Q^{(2)}_{pqr}(t)\\
&+3V^{(3)}_{ijk}(t){J_2}_{ip}(t){J_1}_{jq}(t){J_1}_{kr}(t)Q^{(3)}_{pqr}(t)
+\phantom{3}V^{(3)}_{ijk}(t){J_1}_{ip}(t){J_1}_{jq}(t){J_1}_{kr}(t)
Q^{(4)}_{pqr}(t)\\
&+\phantom{3}V^{(3)}_{ijk}(t)A^-_{pi}(t)A^-_{qj}(t)A^-_{rk}(t)Q^{(5)}_{pqr}(t) 
 +3V^{(3)}_{ijk}(t)A^-_{pi}(t)A^-_{qj}(t)B^-_{rk}(t)Q^{(6)}_{pqr}(t)\\
&+3V^{(3)}_{ijk}(t)A^-_{pi}(t)B^-_{qj}(t)B^-_{rk}(t)Q^{(7)}_{pqr}(t)
+\phantom{3}V^{(3)}_{ijk}(t)B^-_{pi}(t)B^-_{qj}(t)B^-_{rk}(t)Q^{(8)}_{pqr}(t)
\end{aligned}\right.
\end{equation}
\end{widetext}
with vanishing initial conditions for $Q^{(i)}_{pqr}$ and $I_2^-$
Finally, one must add equations leading to $I_l$ and $I_1$
computation, namely
\begin{equation}
\left\{\begin{aligned}
\dot{I}_1&=V^{(4)}_{ijkl}(t)\mathcal{G}_{ij}(t,t)\mathcal{G}_{kl}(t,t),\\
\dot{I}_l&=V^{(3)}_{ijk}(t)A^-_{lk}(t)\mathcal{G}_{ij}(t,t),
\end{aligned}\right.
\end{equation}
where we have used $\mathcal{G}_{lk}(0,t)=A^-_{lk}(t)$. 
Taking into account equations for $\mathbf{X}(t)$ and $M(t)$, this
gives rise to a total of $8f^3+4f^2+7f+3$ equations, that is $97$ for a 2D
system. 

In practice, having found a periodic orbit and for a given $t_0$ along
this orbit, the
coefficient $C_1(T,t_0)$ is computed in two steps:
\begin{itemize}
\item[$\bullet$] Achieve the SVD decomposition of the matrix
  $\mathcal{A}$, appearing on the left-hand side of Eq.~(\ref{LSAB}),
  and compute the projector matrix appearing on the right-hand side
  of Eq.~(\ref{Mproj}).
\item[$\bullet$] Integrate the differential set~(\ref{DSETTRK}) along
  the periodic orbit (starting at point depicted by $t_0$). At any
  time $t$, use the preceding SVD decomposition to obtain a solution
  $\mathcal{X}_0$ and the projector matrix to get the true solution
  $(A_-(t),B_-(t))$ and thus $(A_+(t), B_+(t))$, using Eq.~(\ref{Mproj}).
\end{itemize}

Finally, the coefficient $C_1(T,t_0)$, being a smooth function of $t_0$, 
the average over time $t_0$, leading to the $\hbar$ correction term
$C_1(T)$, can be handled by any conventional integrator.

\section{Trace of the Green's function $G(E)$}
\label{Sec:TG}

Steps leading to the semiclassical contribution $G_l(E)$ from a given
periodic orbit to the trace
of the Green's function $G(E)$ are identical to those giving the
$G_l(\mathbf{q},\mathbf{q}_0,E)$ expression, so that $G_l(E)$ reads
\begin{multline}
\label{trgreen}
G_l(E)=\frac{1}{i\hbar}\frac{T_0}{\left|\det{(m(T_0)-\openone)}\right|^{1/2}}
\exp{\left[\frac{i}{\hbar}S_l(E)-i\frac{\pi}{2}\mu_l\right]} \\
\times\Bigl\{1+i\hbar\left[C_1(T_0)+C_1^{T\rightarrow E}(T_0)\right]
+\mathcal{O}(\hbar^2)\Bigr\},
\end{multline}
where $C_1^{T\rightarrow E}(T)$ is given by
\begin{multline}
C_1^{T\rightarrow E}(T_0)=\frac{1}{2W_l^{(2)}}
\left[\left(C_0^{(1)}\right)^2+C_0^{(2)}\right]\\
-\frac{W_l^{(3)}C_0^{(1)}}{2\left(W_l^{(2)}\right)^2}
-\frac{W_l^{(4)}}{8\left(W_l^{(2)}\right)^2}+
\frac{5}{24}\frac{\left(W_l^{(3)}\right)^2}{\left(W_l^{(2)}\right)^3}.
\end{multline}
$W_l^{(i)}$ (respectively  $C_0^{(i)}$) are the Taylor coefficients of
the $W_l(T)$ (resppectively $C_0(T)$) expansion around $T_0$.

Computation of $W_l^{(i)}$ is much the same as in the Green's function
case, because the functional relation
\begin{equation}
\frac{\partial W_l(T)}{\partial T}=-E(T)
\end{equation}
still holds for a given periodic orbit, $E(T)$ being its energy as
function of its period, which is still given by the value of the
Hamiltonian $H$ taken at any point on 
the corresponding phase space trajectory 
$\mathbf{X}(t,T)=\bigl(\mathbf{q}(t,T),\mathbf{p}(t,T)\bigr)$. Thus,
the Taylor expansion of $\mathbf{X}(t,T)$ around the periodic orbit
$\mathbf{X}(t,T_0)$, will lead to the same expressions for $W_l^{(i)}$
coefficients~(Eq.~\ref{dwdt}) and for $\mathbf{X}^{(n)}(t)$
equations~(Eq.~\ref{dXdt}). The only differences with the preceding
section arise from the 
boundary conditions fulfilled by $\mathbf{X}^{(n)}(t)$, deduced from
the equation $\mathbf{X}(0,T)=\mathbf{X}(T,T)$, i.e., $\mathbf{X}(t,T)$ is
a periodic orbit of period $T$. The Taylor expansion of this relation
leads to the following conditions:
\begin{equation}
\label{BCXP}
\left\{\begin{aligned}
\mathbf{X}^{(1)}(0)&=\mathbf{X}^{(1)}(T_0)+\dot{\mathbf{X}}^{(0)}(T_0),\\
\mathbf{X}^{(2)}(0)&=\mathbf{X}^{(2)}(T_0)+\ddot{\mathbf{X}}^{(0)}(T_0)+
2\dot{\mathbf{X}}^{(1)}(T_0),\\
\mathbf{X}^{(3)}(0)&=\mathbf{X}^{(3)}(T_0)+\dddot{\mathbf{X}}^{(0)}(T_0)+
3\ddot{\mathbf{X}}^{(1)}(T_0)+3\dot{\mathbf{X}}^{(2)}(T_0).
\end{aligned}\right.
\end{equation}
Solutions of the differential set~(\ref{dXdt}) still have the
following formal expressions~(\ref{deltaXt}),
which, inserted in the boundary conditions~(\ref{BCXP}), leads to
equations on $\mathbf{X}^{(i)}(0)$ only
\begin{equation}
\left\{\begin{aligned}
\bigl(\openone-M(T_0)\bigr)\mathbf{X}^{(1)}(0)&=\dot{\mathbf{X}}^{(0)}(T_0),\\
\bigl(\openone-M(T_0)\bigr)\mathbf{X}^{(2)}(0)&=\ddot{\mathbf{X}}^{(0)}(T_0)
+2\dot{\mathbf{X}}^{(1)}(T_0)+\mathbf{F}^{(2)}(T_0),\\
\bigl(\openone-M(T_0)\bigr)\mathbf{X}^{(3)}(0)&=\dddot{\mathbf{X}}^{(0)}(T_0)
+3\ddot{\mathbf{X}}^{(1)}(T_0)\\
&+3\ddot{\mathbf{X}}^{(2)}(T_0)
+\mathbf{F}^{(3)}(T_0).
\end{aligned}\right.
\end{equation}
The matrix $\openone-M(T_0)$ being singular, solving the preceding
linear equations need additional discussion, which, for simplicity, will
focus on $\mathbf{X}^{(1)}(0)$ only. First, the nullspace of 
\mbox{$\openone-M(T_0)^{\top}$} is spanned by 
$\Sigma\dot{\mathbf{X}}^{(0)}(T_0)$, which is obviously orthogonal to 
$\dot{\mathbf{X}}^{(0)}(T_0)$, the right hand-side of the equation for
$\mathbf{X}^{(1)}(0)$, thus showing that this equation admits
solutions. Then, the 
nullspace of \mbox{$\openone-M(T_0)$} being spanned by
$\dot{\mathbf{X}}^{(0)}(T_0)$, the whole set of solutions reads
\begin{equation}
\mathbf{X}^{(1)}(0)=\mathbf{X}^{(1)}_0(0)+\alpha\dot{\mathbf{X}}^{(0)}(T_0),
\end{equation}
where $\mathbf{X}^{(1)}_0(0)$ is a particular solution of the
equation. Actually, the term $\alpha\dot{\mathbf{X}}^{(0)}(T_0)$
corresponds to an displacement of the initial conditions along the
flow, which, of course, gives back the same periodic orbit (at first order
in $T-T_0$). We thus expect that this term has a vanishing
contribution to $W_l^{(2)}$, which is easily verified when
inserting the general solution in the $W_l^{(2)}$ expression (taken at
time $t=T_0$):
\begin{equation}
\begin{aligned}
W_l^{(2)}&=-\left(\mathbf{X}^{(1)}_0(0)-\dot{\mathbf{X}}^{(0)}(T_0)\right.\\
&\phantom{=-\left(\mathbf{X}^{(1)}_0(0)\right.}
\left.+\alpha\dot{\mathbf{X}}^{(0)}(T_0)\right)\cdot
\bm{\nabla} H\left(\mathbf{X}^{(0)}(T_0)\right) \\
&=-\mathbf{X}^{(1)}_0(0)\cdot\bm{\nabla} H\left(\mathbf{X}^{(0)}(T_0)\right)
\end{aligned}
\end{equation}
because of the Hamilton's equations
$\dot{\mathbf{X}}^{(0)}(T_0)=\Sigma\bm{\nabla} 
H\left(\mathbf{X}^{(0)}(T_0)\right)$.

These two properties also hold in the cases of $\mathbf{X}^{(2)}(0)$
and $\mathbf{X}^{(3)}(0)$, but are slightly more complicated to
establish because the right-hand sides of the equations involve
$\mathbf{F}^{(i)}(T_0)$ and derivatives of $\mathbf{X}^{(i)}(T_0)$.

Thus, integrating the same differential sets that were used for
$G(\mathbf{q},\mathbf{q}_0,E)$, one is able to compute the first four
derivatives of the action, $W_l^{(i)}$, with respect to the period.

Starting from the $C_0^{(0)}(T)$ expression
\begin{equation}
C_0^{(0)}(T)=\ln\,{T}-\frac12\ln{|\partial_ET|}
-\frac12\ln{|\det{(m(T)-\openone)}|}
\end{equation}
and using the fact that
$\partial_ET=1/\partial_TE=-1/\partial_T^2W_l$, one obtains
\begin{equation}
\label{dc0dttr}
\left\{\begin{aligned}
C_0^{(1)}(T_0)&=\frac{1}{T_0}+\frac{1}{2}\frac{W_l^{(3)}}{W_l^{(2)}}
-\frac12\frac{d}{dT}\ln{|\det{(m(T)-\openone)}|}, \\
C_0^{(2)}(T_0)&=-\frac{1}{T_0^2}+\frac{1}{2}\frac{W_l^{(4)}}{W_l^{(2)}}
-\frac{1}{2}\left(\frac{W_l^{(3)}}{W_l^{(2)}}\right)^{2}\\
&\phantom{=-\frac{1}{T_0^2}}
-\frac12\frac{d^2}{dT^2}\ln{|\det{(m(T)-\openone)}|},
\end{aligned}\right.
\end{equation}
which means that one is left with the calculation of derivatives of
$\ln{|\det{(m(T)-\openone)}|}$ with respect to the period $T$. As shown
in the Appendix, $\det{(m(T)-\openone)}$ is given by the
determinant of the $2f\times 2f$ matrix $N(T)$ defined as follows:
\begin{equation}
N(T)=M(T)-\left(\openone-\mathcal{P}_{\parallel}(T)
-\mathcal{P}_{\perp}(T)\right),
\end{equation}
where we have introduced $\mathcal{P}_{\parallel}(T)$ (respectively
$\mathcal{P}_{\perp}(T)$) the projector on the direction parallel to the
flow (respectively perpendicular to the energy shell), more precisely,
the $\mathcal{P}_{\parallel}(T)$ and $\mathcal{P}_{\perp}(T)$ expressions are
\begin{equation}
\mathcal{P}_{\parallel}=\mathbf{e}_{\parallel}\cdot
\mathbf{e}_{\parallel}^{\top}\quad\text{and}\quad
\mathcal{P}_{\perp}=\mathbf{e}_{\perp}\cdot
\mathbf{e}_{\perp}^{\top}=-\Sigma\mathcal{P}_{\parallel}\Sigma,
\end{equation}
where $\mathbf{e}_{\parallel}$ is the unit vector tangent to the flow
at initial (and thus final) time and 
$\mathbf{e}_{\perp}=\Sigma\mathbf{e}_{\parallel}$.
Now, using again formula~(\ref{dlndet}), derivatives of
$\det{(m(T)-\openone)}$ with respect to the period read
\begin{widetext}
\begin{equation}
\label{dndt}
\left\{\begin{aligned}
\frac{d}{dT}\left(\det{(m(T)-\openone)}\right) &=
\mathrm{tr}\left(N(T_0)^{-1}\frac{dN(T_0)}{dT}\right),\\
\frac{d^2}{dT^2}\left(\det{(m(T)-\openone)}\right) &=\mathrm{tr}
\left(N^{-1}(T_0)
\frac{d^2N(T_0)}{dT^2}-N(T_0)^{-1}\frac{dN(T_0)}{dT}
N(T_0)^{-1}\frac{dN(T_0)}{dT}\right)
\end{aligned}\right.
\end{equation}
with
\begin{equation}
\left\{\begin{aligned}
\frac{dN(T_0)}{dT}&=\frac{dM(T_0)}{dT}+\frac{d\mathcal{P}_{\parallel}(T_0)}{dT}
-\Sigma\frac{d\mathcal{P}_{\parallel}(T_0)}{dT}\Sigma,\\
\frac{d^2N(T_0)}{dT^2}&=\frac{d^2M(T_0)}{dT^2}
+\frac{d^2\mathcal{P}_{\parallel}(T_0)}{dT^2}
-\Sigma\frac{d^2\mathcal{P}_{\parallel}(T_0)}{dT^2}\Sigma.
\end{aligned}\right.
\end{equation}
As seen previously (Sec.~\ref{C1TE}), $\frac{dM(T_0)}{dT}$ and
$\frac{d^2M(T_0)}{dT^2}$ are expressed 
in terms of the coefficients $M^{(i)}(t)$ of the Taylor expansion of
the monodromy matrix $M(t,T)$ 
(associated with the periodic orbit $\mathbf{X}(t,T)$ of period $T$)
around the periodic orbit $\mathbf{X}^{(0)}(t)$ of period $T_0$, see
Eq.~(\ref{dmdt}). 

Inserting the Taylor expansion of $\dot{\mathbf{X}}(T)$ around $T_0$ in
the $\mathcal{P}_{\parallel}(T)$ expression, namely,
\begin{equation}
\label{pparT}
\mathcal{P}_{\parallel}(T)=\frac{1}{\|\dot{\mathbf{X}}(T)\|^2}
\dot{\mathbf{X}}(T)\cdot\dot{\mathbf{X}}(T)^{\top},
\end{equation}
one obtains the derivatives of $\mathcal{P}_{\parallel}(T)$ with
respect to $T$:
\begin{equation}
\left\{\begin{aligned}
\frac{d\mathcal{P}_{\parallel}(T_0)}{dT}&=
\frac{1}{\|\dot{\mathbf{X}}^{(0)}\|^2}
\left(\dot{\mathbf{X}}^{(1)}\cdot\dot{\mathbf{X}}^{{(0)}^{\top}}+
\dot{\mathbf{X}}^{(0)}\cdot\dot{\mathbf{X}}^{(1)^{\top}}\right)
-2\frac{\dot{\mathbf{X}}^{(0)^{\top}}\!\!\cdot\dot{\mathbf{X}}^{(1)}}
{\|\dot{\mathbf{X}}^{(0)}\|^2},
\mathcal{P}_{\parallel}(T_0) \\
\frac{d^2\mathcal{P}_{\parallel}(T_0)}{dT^2}&=
\frac{1}{\|\dot{\mathbf{X}}^{(0)}\|^2}
\left(\dot{\mathbf{X}}^{(2)}\cdot\dot{\mathbf{X}}^{{(0)}^{\top}}+
\dot{\mathbf{X}}^{(0)}\cdot\dot{\mathbf{X}}^{(2)^{\top}}
+2\dot{\mathbf{X}}^{(1)}\cdot\dot{\mathbf{X}}^{{(1)}^{\top}}\right) \\
&+\left(
8\frac{\left(\dot{\mathbf{X}}^{(0)^{\top}}
\!\!\cdot\dot{\mathbf{X}}^{(1)}\right)^2}
{\|\dot{\mathbf{X}}^{(0)}\|^4}
-2\frac{\dot{\mathbf{X}}^{(0)^{\top}}\!\!\cdot\dot{\mathbf{X}}^{(2)}}
{\|\dot{\mathbf{X}}^{(0)}\|^2}
-2\frac{\dot{\mathbf{X}}^{(1)^{\top}}\!\!\cdot\dot{\mathbf{X}}^{(1)}}
{\|\dot{\mathbf{X}}^{(0)}\|^2}\right)\mathcal{P}_{\parallel}(T_0)\\
&-4\frac{\dot{\mathbf{X}}^{(0)^{\top}}\!\!\cdot\dot{\mathbf{X}}^{(1)}}
{\|\dot{\mathbf{X}}^{(0)}\|^4}
\left(\dot{\mathbf{X}}^{(1)}\cdot\dot{\mathbf{X}}^{{(0)}^{\top}}+
\dot{\mathbf{X}}^{(0)}\cdot\dot{\mathbf{X}}^{(1)^{\top}}\right),
\end{aligned}\right.
\end{equation}
\end{widetext}
where all $\dot{\mathbf{X}}^{(i)}$ are evaluated at time
$t=0$.

Gathering the preceding expressions into Eq.~(\ref{dndt})
allows us to compute $\ln{\det{(m(T)-\openone)}}$ derivatives,
which, inserted together with derivatives of the action, in
Eq.~(\ref{dc0dttr}) gives the numerical values for $C_0^{(1)}(T_0)$ and
$C_0^{(2)}(T_0)$, which finally leads to the additional $\hbar$
correction $C_1^{T\rightarrow E}(T_0)$.

\section{Application to the 2D hydrogen atom in a magnetic field}
\label{bidihy}
The hydrogen atom is one example of quantum system whose classical
counterpart depicts a chaotic behavior and has been widely studied (see
for, e.g., Ref.~\cite{FW89} for a complete review). It has now
become a very useful tool for testing new ideas and tools in the
quantum chaos area, both on the semiclassical~\cite{M99,MD00} or
universality~\cite{JGD98} points of view, especially because computing
very highly excited states has become a standard task on a regular
workstation, allowing the semiclassical regime to be reached easily.
Even if one would have preferred to work with the real
hydrogen atom (i.e., the three-dimensional one), in this paper we will focus 
on the two dimensional hydrogen atom in a magnetic field, because
taking into account invariance by rotation around the magnetic field, 
gives rise to centrifugal terms in the Hamiltonian (typically
$\mathbf{L}^2\hbar^2/2r^2$) which would also contribute to $\hbar$
corrections and would need a study on its own. One must also notice
that, even if the classical 
dynamics are identical for both cases, the fact that the magnetic
field axis is no longer a rotation axis in the 2D case gives rise to
slight modifications in the Maslov indices~\cite{EW91,MD00,B89}.

\subsection{Quantum and classical properties}

In atomic units the Hamiltonian of the 2D hydrogen in magnetic field
reads
\begin{equation}
H=\frac{1}{2} \mathbf{p}^2-\frac{1}{\sqrt{x^2+y^2}}+\frac{1}{8}\gamma^2y^2,
\end{equation}
where $\gamma=B/B_0$, with $B_0=2.35\times10^5 T$.
The classical counterpart of this Hamiltonian has a scaling property,
that is, if we define new variables by
\begin{equation}
\label{scaling}
  \left\{\begin{array}{l}
\tilde{\mathbf{r}}=\gamma^{2/3}\mathbf{r}, \\
\tilde{\mathbf{p}}=\gamma^{-1/3}\mathbf{p}, \\
\tilde{t}=\gamma t, \\
\end{array}\right.
\end{equation}
we obtain a new Hamiltonian $\tilde{H}$ given by
\begin{equation}
  \tilde{H}=\gamma^{-2/3}H=\frac {{\tilde{\mathbf{p}}}^2}2
  -\frac 1{\sqrt{\tilde{x}^2+\tilde{y}^2}}+\frac{\tilde{y}^2}8,
\end{equation}
which does not depend on $\gamma$ anymore. The classical dynamics of
this Hamiltonian is entirely fixed by the scaled energy $\epsilon$
given by:
\begin{equation}
  \label{eet}
  \epsilon=\gamma^{-2/3}E.
\end{equation}
All properties of the classical trajectories of the original
Hamiltonian can be deduced from the scaled dynamics using the scaling
transformation~\eqref{scaling}. From the quantum point of view, this
scaling introduces an effective 
$\hbar$ value, which is easily seen on the scaled Schr{\"o}dinger
equation, $\tilde{H}\psi=\epsilon\psi$, for a fixed scaled energy
$\epsilon$:
\begin{equation}
  \biggl[-\frac{\gamma^{2/3}}2\Delta_{\tilde{\mathbf{r}}}
  -\frac{1}{\sqrt{\tilde{x}^2+\tilde{y}^2}}+
  \frac{\tilde{y}^2}8\biggr]\psi=\epsilon\psi.
\end{equation}
Thus, the effective $\hbar$ is given by $\gamma^{1/3}$ and so at
a fixed value of the scaled energy $\epsilon$, the semiclassical limit
is obtained  when $\gamma$ tends to 0.

The singularity in the classical equations of motion due to the
divergence of the Coulomb potential at $\mathbf{r}=\mathbf{0}$ is regularized
using the semiparabolic coordinates
$(u=\sqrt{\tilde{r}+\tilde{x}},\,v=\sqrt{\tilde{r}-\tilde{x}})$,
giving rise to the following effective 
classical Hamiltonian~\cite{FW89,Englefield}:
\begin{equation}
  \mathcal{H}=\frac 12 p_{u}^2+\frac 12 p_{v}^2-\epsilon(u^2+v^2)+\frac
  18u^2v^2(u^2+v^2),
\end{equation}
the trajectories
corresponding to the original problem are obtained when fixing total
energy $\mathcal{H}=2$. The associated quantum Hamiltonian reads
\begin{equation}
  \hat{\mathcal{H}}(\hbar)=-\frac{\hbar^2}{2}
\left(\frac{\partial^2}{\partial u^2}+
\frac{\partial^2}{\partial v^2}\right)-\epsilon(u^2+v^2)+\frac
  18u^2v^2(u^2+v^2),
\end{equation}
which separates into kinetic and potential energy, so that the
semiclassical formula derived in the preceding sections applied to
the associated quantum Green's function
$G(z,\hbar)$, the hydrogen in a magnetic field being recovered for $z=2$
(actually $z/2$ corresponds to the nucleus charge)
\begin{equation}
G(z,\hbar)=\frac{1}{z-\hat{\mathcal{H}}(\hbar)}=
\sum_{\tau}\frac{|\tau,\hbar\rangle\langle\tau,\hbar|}
{z-\lambda_{\tau}(\hbar)},
\end{equation}
where $|\tau,\hbar\rangle$ is an (normalized) eigenvector of
$\hat{\mathcal{H}}(\hbar)$ for the eigenenergy $\lambda_{\tau}(\hbar)$,
$\tau$ representing the set of quantum labels, i.e., level number and
symmetry properties (see below), describing $|\tau,\hbar\rangle$.
The matrix element 
$\langle\mathbf{q}|G(z,\hbar)|\mathbf{q}_0\rangle$, where 
$\mathbf{q}=(u,v)$ then reads
\begin{equation}
\langle\mathbf{q}|G(z,\hbar)|\mathbf{q}_0\rangle=
\sum_{\tau}\psi_{\tau,\hbar}(\mathbf{q})\psi_{\tau,\hbar}(\mathbf{q}_0)
\frac{1}{z-\lambda_{\tau}(\hbar)},
\end{equation}
where 
$\psi_{\tau,\hbar}(\mathbf{q})=\langle\mathbf{q}|\tau,\hbar\rangle$
has been supposed to be real,  with
$\hat{\mathcal{H}}(\hbar)$ being invariant under
$\mathbf{p}\rightarrow-\mathbf{p}$.
Taking $z=\lambda$ on the real axis, the
imaginary part of $\langle\mathbf{q}|G(z,\hbar)|\mathbf{q}_0\rangle$, becomes
\begin{equation}
\label{imgqq0}
-\frac{1}{\pi}\mathrm{Im}\,\langle\mathbf{q}|G(\lambda,\hbar)
|\mathbf{q}_0\rangle
=\sum_{\tau}\psi_{\tau,\hbar}(\mathbf{q})\psi_{\tau,\hbar}(\mathbf{q}_0)
\delta\left(\lambda-\lambda_{\tau}(\hbar)\right)
\end{equation}
to which any classical path going from $\mathbf{q}$ to
$\mathbf{q}_0$ at energy $\lambda$, gives the following contribution
(see Eq.~(\ref{gqqofin})):
\begin{multline}
\label{imgqq0sc}
-\frac{1}{\pi}\mathrm{Im}\,\langle\mathbf{q}|G(\lambda,\hbar)
|\mathbf{q}_0\rangle_l=\frac{2}{(2\pi\hbar)^{3/2}}
\mathcal{A}_l\left\{\cos{\left(\frac
        1{\hbar}S_l+\phi_l\right)}\right.\\
\left.-\hbar\mathcal{C}_l
    \,\sin{\left(\frac1{\hbar}S_l+\phi_l\right)}\right\}
\end{multline}
provided it is far enough from any bifurcation and that $\mathbf{q}$ and
$\mathbf{q}_0$ are not conjugate points for this trajectory. Amplitudes
and phases being defined by
\begin{equation}
\label{CPGQQ0}
\left\{\begin{aligned}
\mathcal{A}_l &=\frac{1}{\left|W_l^{(2)}\,\det{\,J_1(T_0)}\right|^{1/2}},\\
S_l &= S(\mathbf{q},\mathbf{q}_0,\lambda), \\
\phi_l &= -\frac{\pi}{2}(\tilde{\nu}_l+\frac{1}{2}),\\
\mathcal{C}_l &= C_1(\mathbf{q},\mathbf{q}_0,T_0)
+C_1^{T\rightarrow E}(\mathbf{q},\mathbf{q}_0,T_0).
\end{aligned}\right.
\end{equation}

Neglecting $\hbar$ corrections in Eq.~(\ref{imgqq0sc}),
the Fourier transform with respect to the variable
$\zeta=1/\hbar$ of the following function:
\begin{multline}
\label{scaled0}
g_0(\zeta)=\frac{(2\pi)^{3/2}}{2\zeta^{3/2}}\times-\frac{1}{\pi}
\mathrm{Im}\,\langle\mathbf{q}|G(\lambda,\zeta)
|\mathbf{q}_0\rangle=\\
\frac{(2\pi)^{3/2}}{2}
\sum_{\tau}\psi_{\tau,\zeta}(\mathbf{q})\psi_{\tau,\zeta}(\mathbf{q}_0)
\zeta^{-3/2}\delta\left(\lambda-\lambda_{\tau}(\zeta)\right)
\end{multline}
will depict peaks at the classical actions $S_l/2\pi$, with complex
amplitude $\mathcal{A}_l\,\exp{i\phi_l}/2$, which has been extensively
used to compare the exact quantum Green's function with its semiclassical
estimation at the leading order in $\hbar$. In the same way, the
Fourier transform of the following function:
\begin{multline}
\label{scaled1}
g_1(\zeta)=-\frac{(2\pi)^{3/2}}{2}\sum_{\tau}
\psi_{\tau,\zeta}
(\mathbf{q})\psi_{\tau,\zeta}(\mathbf{q}_0)\zeta^{-1/2}
\delta\left(\lambda-\lambda_{\tau}(\zeta)\right)\\
-\zeta\sum_l\mathcal{A}_l\,\cos{\left(
        \zeta S_l+\phi_l\right)}
\end{multline}
will also depict peaks at the classical actions $S_l/2\pi$, whose complex
amplitude, given by
\begin{equation}
\frac{1}{2i}\mathcal{A}_l\mathcal{C}_l\exp{i\phi_l}
\end{equation}
allows us to extract the numerical value of the $\hbar$ correction $C_l$.

The energy $\lambda$ being fixed, the 
$\delta(\lambda-\lambda_{\tau}(\zeta))$ function selects the values
$\zeta_{\tau}(\lambda)$ of $\zeta$ for which $\lambda$ is an
eigenvalue, transforming Eqs.~(\ref{scaled0}) and~(\ref{scaled1}) into
\begin{widetext}
\begin{equation}
\label{scaled3}
\left\{\begin{aligned}
g_0(\zeta)&=\phantom{-}\frac{(2\pi)^{3/2}}{4}\sum_{\tau}
\frac{\psi_{\tau,\zeta}(\mathbf{q})\psi_{\tau,\zeta}(\mathbf{q}_0)}
{\langle\tau,\zeta|\frac{\mathbf{p}^2}{2}|\tau,\zeta\rangle}
\zeta^{3/2}
\delta\left(\zeta-\zeta_{\tau}(\lambda)\right), \\
g_1(\zeta)&=-\frac{(2\pi)^{3/2}}{4}\sum_{\tau}
\frac{\psi_{\tau,\zeta}(\mathbf{q})\psi_{\tau,\zeta}(\mathbf{q}_0)}
{\langle\tau,\zeta|\frac{\mathbf{p}^2}{2}|\tau,\zeta\rangle}
\zeta^{5/2}\delta\left(\zeta-\zeta_{\tau}(\lambda)\right)
-\zeta\sum_l\mathcal{A}_l\,\cos{\left(\zeta S_l+\phi_l\right)}.
\end{aligned}\right.
\end{equation}
\end{widetext}

Stepping to the case of the trace of the Green's function, the
preceding relations~(\ref{imgqq0}) and~(\ref{imgqq0sc}) becomes
\begin{equation}
-\frac{1}{\pi}\mathrm{Im}\,\,\mathrm{Tr}\,G(\lambda,\hbar)
=\sum_{\tau}\delta\left(\lambda-\lambda_{\tau}(\hbar)\right)
\end{equation}
and, see Eq.~(\ref{trgreen}):
\begin{multline}
-\frac{1}{\pi}\mathrm{Im}\,\,\mathrm{Tr}\,G(\lambda,\hbar)_l=
-\frac{1}{\pi\hbar}
\mathcal{A}_l^{\mathrm{tr}}\left\{\cos{\left(\frac
        1{\hbar}S_l^{\mathrm{tr}}+\phi_l^{\mathrm{tr}}\right)}\right.\\
\left.-\hbar\mathcal{C}_l^{\mathrm{tr}}
    \,\sin{\left(\frac1{\hbar}S_l^{\mathrm{tr}}
+\phi_l^{\mathrm{tr}}\right)}\right\},
\end{multline}
where $S_l^{\mathrm{tr}}$ is the action of the periodic orbit and
\begin{equation}
\label{CPTRG}
\left\{\begin{aligned}
\mathcal{A}_l^{\mathrm{tr}} &=\frac{T_0}{\left|\det{(m(T_0)-\openone)}
\right|^{1/2}},\\
\phi_l^{\mathrm{tr}} &= -\frac{\pi}{2}\mu_l,\\
\mathcal{C}_l^{\mathrm{tr}} &= C_1(T_0)+C_1^{T\rightarrow E}(T_0),
\end{aligned}\right.
\end{equation}
so that the classical quantities $S_l^{\mathrm{tr}}$, 
$\mathcal{A}_l^{\mathrm{tr}}$, and the $\hbar$ correction
$\mathcal{C}_l^{\mathrm{tr}}$ can be obtained by taking the Fourier
transform of the following expressions with respect to the variable
$\zeta$:
\begin{widetext}
\begin{equation}
\label{scaledtr}
\left\{\begin{aligned}
g_0^{\mathrm{tr}}(\zeta)&=\phantom{-}\frac{\pi}{2}\sum_{\tau}
\frac{1}
{\langle\tau,\zeta|\frac{\mathbf{p}^2}{2}|\tau,\zeta\rangle}
\zeta^{2}\delta\left(\zeta-\zeta_{\tau}(\lambda)\right),\\
g_1^{\mathrm{tr}}(\zeta)&=-\frac{\pi}{2}\sum_{\tau}
\frac{1}
{\langle\tau,\zeta|\frac{\mathbf{p}^2}{2}|\tau,\zeta\rangle}
\zeta^{3}\delta\left(\zeta-\zeta_{\tau}(\lambda)\right)
-\zeta\sum_l\mathcal{A}_l^{\mathrm{tr}}
\,\cos{\left(\zeta S_l^{\mathrm{tr}}+\phi_l^{\mathrm{tr}}\right)}.
\end{aligned}\right.
\end{equation}
\end{widetext}

\subsection{Computing quantum quantities}

Focusing on the $\lambda=2$ value, the 2D hydrogen in a magnetic field case,
one has to find effective $\hbar$ values for which $2$ is an
eigenvalue of the Schr{\"o}dinger equation 
$\hat{\mathcal{H}}(\hbar)\psi(u,v)=2\psi(u,v)$,
which is conveniently written as follows:
\begin{multline}
\left[2+\epsilon(u^2+v^2)-\frac
  18u^2v^2(u^2+v^2)\right]\psi(u,v)\\
=\hbar^2\left[-\frac{1}{2}\left(\frac{\partial^2}{\partial u^2}+
\frac{\partial^2}{\partial v^2}\right)\right]\psi(u,v)
\end{multline}
such that $\sigma=\hbar^2$ appears to be a solution of a generalized eigenvalue
problem $(A-\sigma B)\psi=0$, with

\begin{equation}
\left\{\begin{aligned}
A&=2+\epsilon(u^2+v^2)-\frac{1}{8}u^2v^2(u^2+v^2), \\
B&=-\frac{1}{2}
\left(\frac{\partial^2}{\partial u^2}+
\frac{\partial^2}{\partial v^2}\right).
\end{aligned}\right.
\end{equation}
The preceding operators $A$, $B$, and thus $\hat{\mathcal{H}}(\hbar)$
are invariant under all transformations 
belonging to the symmetry group $C_{4v}$, leading to four
nondegenerate series of energy levels, labeled EEE, EEO, OOE
and OOO according to Ref.~\cite{MWR89} and a twofold degenerate series
EO and OE, where E means even and O means odd, the first two letters
referring to the $u\rightarrow-u$ and $v\rightarrow-v$ symmetries, the
third letter to $u\leftrightarrow v$. Actually, because of the
definition of the semiparabolic coordinates $(u,v)$, only
eigenvectors invariant under the parity symmetry
$\psi(-u,-v)=\psi(u,v)$ correspond to eigenvectors of the 2D hydrogen in
a magnetic field, allowing us, in principle, to drop the OE and EO
series~\cite{FW89,Englefield}. However, from the semiclassical point of
view, one would have 
to extend all preceding sections to symmetry-projected propagator and
Green's function~\cite{R89}, and thus to take into account
symmetry properties of the classical Green's function, which is beyond
the scope of this paper. For this reason, we also include the OE and
EO series in the remainder of this paper.

Finally, eigenvalues and eigenvectors are obtained by solving the
matrix representation of the generalized eigenvalue problem 
$(A-\sigma B)\psi=0$ in sturmian bases (one for each symmetry
class)~\cite{FW89}, using the Lanczos algorithm. Typically, we have
computed effective $\hbar$ values ranging from $0$ to $124$, which for
scaled energy $\epsilon=-0.1$ corresponds to roughly $61\,000$
eigenvalues in total. One must notice that the generalized eigenvectors 
$|\widetilde{\tau,\hbar}\rangle$, for a fixed $\hbar$ value, are actually
orthogonal for the scalar product defined by operator $B=\mathbf{p}^2/2$:
\begin{equation}
\langle\widetilde{\tau,\hbar}|\frac{\mathbf{p}^2}{2}
|\widetilde{\tau',\hbar}\rangle=\delta_{\tau\tau'}
\end{equation}
so that the $|\widetilde{\tau,\hbar}\rangle$ and $|\tau,\hbar\rangle$
relations read
\begin{equation}
\left\{\begin{aligned}
|\tau,\hbar\rangle&=\frac{1}{\sqrt{\langle\widetilde{\tau,\hbar}
|\widetilde{\tau,\hbar}\rangle}}|\widetilde{\tau,\hbar}\rangle,\\
|\widetilde{\tau,\hbar}\rangle&=\frac{1}
{\sqrt{\langle\tau,\hbar|B|\tau,\hbar\rangle}}|\tau,\hbar\rangle,
\end{aligned}\right.
\end{equation}
giving rise to $g_{0,1}(\zeta)$~(\ref{scaled3}) and  
$g_{0,1}^{\mathrm{tr}}(\zeta)$~(\ref{scaledtr}) expressions in terms of the
computed eigenvectors:
\begin{widetext}
\begin{equation}
\label{scaleddef}
\left\{\begin{aligned}
g_0(\zeta)&=\phantom{-}\frac{(2\pi)^{3/2}}{4}\sum_{\tau}
\tilde{\psi}_{\tau,\zeta}(\mathbf{q})\tilde{\psi}_{\tau,\zeta}(\mathbf{q}_0)
\,\zeta^{3/2}\,\delta\left(\zeta-\zeta_{\tau}(2)\right),\\
g_1(\zeta)&=-\frac{(2\pi)^{3/2}}{4}\sum_{\tau}
\tilde{\psi}_{\tau,\zeta}(\mathbf{q})\tilde{\psi}_{\tau,\zeta}(\mathbf{q}_0)
\,\zeta^{5/2}\,\delta\left(\zeta-\zeta_{\tau}(2)\right)
-\zeta\sum_l\mathcal{A}_l\,\cos{\left(\zeta S_l+\phi_l\right)}, \\
g_0^{\mathrm{tr}}(\zeta)&=\phantom{-}\frac{\pi}{2}\sum_{\tau}
\langle\widetilde{\tau,\zeta}
|\widetilde{\tau,\zeta}\rangle
\,\zeta^{2}\,\delta\left(\zeta-\zeta_{\tau}(2)\right),\\
g_1^{\mathrm{tr}}(\zeta)&=-\frac{\pi}{2}\sum_{\tau}
\langle\widetilde{\tau,\zeta}
|\widetilde{\tau,\zeta}\rangle
\,\zeta^{3}\,\delta\left(\zeta-\zeta_{\tau}(2)\right)
-\zeta\sum_l\mathcal{A}_l^{\mathrm{tr}}
\,\cos{\left(\zeta S_l^{\mathrm{tr}}+\phi_l^{\mathrm{tr}}\right)}.
\end{aligned}\right.
\end{equation}
\end{widetext}

As explained previously, the Fourier transform of the two functions
$g_1$ and $g_1^{\mathrm{tr}}$
will depicts peaks at classical actions and $\hbar$ corrections are
obtained from the amplitude of these peaks. However, in the case of signal
given by $c(t)=\sum a_n\exp{(i\omega_n t)}$, it is now well known
that the harmonic inversion method is very well suited and is much
more powerful than the conventional Fourier 
transform to extract unknown frequencies $\omega_n$ and amplitudes
$a_n$~\cite{M99}. In our case the signals are the two functions
$g_1(\zeta)$ and  
$g_1^{\mathrm{tr}}(\zeta)$, which are of the form
$\sum_l\mathcal{A}_l\mathcal{C}_l\,\sin{\left(\zeta S_l+\phi_l\right)}$
besides contributions from all other types of orbits (ghost,
continuous family, etc.).

\subsection{$\hbar$ corrections for $G(\mathbf{q},\mathbf{q}_0,2)$}

Orbits having initial and final points at the nucleus (i.e., 
$\mathbf{q}=\mathbf{q}_0=\mathbf{0}$) are of special interest because
they are involved in semiclassical estimation of the photoionization
cross-section~\cite{B89,GD92}, which can be directly compared to
experimental results~\cite{Holle88,Main94}. 
Even if the full $\hbar$ expansion of the
cross-section does not reduce to $G(\mathbf{0},\mathbf{0},2)$
contributions, all closed orbits are well known and classified, so that
this case remains a nice example of $\hbar$ corrections for
$G(\mathbf{q},\mathbf{q}_0,2)$. 
\begin{figure}[ht]
\includegraphics[width=7cm,angle=-90]{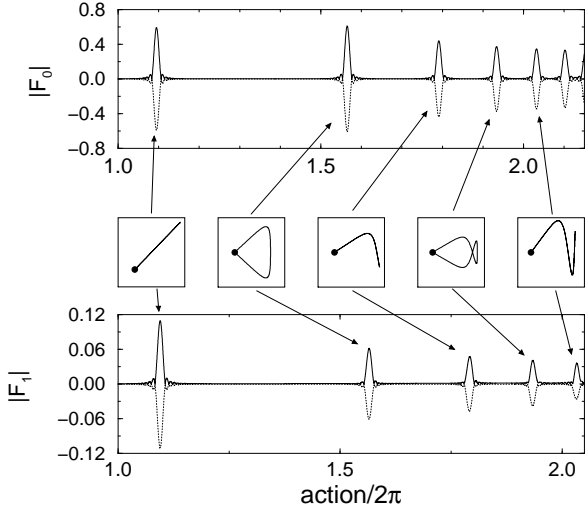}
\caption{\label{G00TF} Modulus of the windowed Fourier transforms
  $F_0$ (solid line, upper plot) and $F_1$ (solid line, lower plot),
  see Eq.~(\ref{WFTGQQ0}), 
  of the quantum functions $g_0$ (leading order in $\hbar$) and $g_1$
  (first order $\hbar$ correction), see
  Eq.~(\ref{scaleddef}), associated with the quantum Green function
$G(\mathbf{q},\mathbf{q}_0,2)$
in the case of the 2D hydrogen atom in a magnetic field
 and for  $\mathbf{q}=\mathbf{q}_0=\mathbf{0}$ (see
 Sec.~\protect\ref{bidihy} 
for all details). As expected from semiclassical
formula~(\protect\ref{gqqofin}), peaks are appearing at action (i.e. 
$\int\mathbf{p}\,d\mathbf{q}/2\pi$) corresponding to classical
orbits having initial and final positions at the nucleus. For the first
five ones, the trajectory in the $(u,v)$ plane are also plotted, the
nucleus being depicted by the black circle. The agreement with the
semiclassical estimations of these functions (dotted lines) is
excellent, even if discrepancies in the amplitude of last two peaks in
the lower plot can be observed. These are actually a manifestation of
limitation of the Fourier transform and not inacurrate calculations of
the $\hbar$ corrections, as it is emphasized by the quantitative
comparison (using harmonic inversion) displayed by
Table~\protect~\ref{tabclosnum}.} 
\end{figure}

\begin{table}
\caption{\label{tabclosed}Classical properties of closed orbits involved in
  the semiclassical expansion of the quantum Green's 
  function $G(\mathbf{q},\mathbf{q}_0,2)$ of the 2D hydrogen atom in
  a magnetic field, for the case
  $\mathbf{q}=\mathbf{q}_0=\mathbf{0}$. Because each closed orbit
  corresponds either to a half-periodic orbit or a periodic orbit, we
  have labeled them with the four-disk code of the corresponding periodic
orbit~\protect\cite{EW90,H95}. Their trajectories in the
  $(u,v)$ plane are shown in Fig.~\protect\ref{G00TF}. $S_l$ is the
  reduced action (i.e., $\int\mathbf{p}\,d\mathbf{q}/2\pi$), $T_l$ is the
period, $\mathcal{A}_l$ is the leading semiclassical
amplitude, $\tilde{\nu}_l$ is the Maslov index,
$\mathcal{C}_l$ is the first order $\hbar$ correction, given by the
sum $C_1(\mathbf{0},\mathbf{0},T_l)+C_1^{T\rightarrow E}(\mathbf{0}
,\mathbf{0},T_l)$, see Eq.~(\protect\ref{CPGQQ0}).}
\begin{ruledtabular}
\begin{tabular}{l@{}c@{}c@{}c@{}c}
\multicolumn{1}{c}{Code} & 
\multicolumn{1}{c}{$S_l$} & 
\multicolumn{1}{c}{$T_l$} & 
\multicolumn{1}{c}{$\mathcal{A}_l$} & 
\multicolumn{1}{c}{$\tilde{\nu}_l$} \\
\hline
$\overline{13}$     & 1.094\,570\,5 & 2.425\,093\,3 & 0.295\,342\,6 & 1  \\
$\overline{1243}$    & 1.564\,998\,2 & 3.600\,137\,4 & 0.152\,365\,0 & 2 \\
$\overline{121343}$   & 1.791\,060\,7 & 4.286\,257\,7 & 0.109\,503\,9 & 3  \\
$\overline{12124343}$  & 1.933\,522\,1 & 4.796\,775\,8 & 0.093\,368\,7 & 4  \\
$\overline{1212134343}$ & 2.031\,948\,2 & 5.214\,323\,3 & 0.086\,142\,0 & 5  \\
& & & & \\
\multicolumn{1}{c}{Code} &
\multicolumn{1}{c}{$C_1(\mathbf{0},\mathbf{0},T_l)$} & 
\multicolumn{1}{c}{$C_1^{T\rightarrow E}(\mathbf{0},\mathbf{0},T_l)$} & 
\multicolumn{1}{c}{$\mathcal{C}_l$}& \\
\hline
$\overline{13}$     & -0.202\,769\,9 & 0.016\,539\,4 & -0.186\,230\,5 &\\
$\overline{1243}$    & -0.119\,409\,3 & 0.019\,741\,2 & -0.099\,668\,1 &\\
$\overline{121343}$   & -0.148\,282\,2 & 0.041\,175\,5 & -0.107\,106\,7 &\\
$\overline{12124343}$  & -0.172\,990\,6 & 0.071\,748\,0 & -0.101\,242\,7 &\\
$\overline{1212134343}$ & -0.192\,904\,3 & 0.117\,464\,5 & -0.075\,439\,8 &
\end{tabular}
\end{ruledtabular}
\end{table}

\begin{table}
\caption{\label{tabclosnum}Numerical comparison between the
  theoretical $\hbar$ corrections $\mathcal{C}_l$ for the quantum
  Green's function $G(\mathbf{q},\mathbf{q}_0,2)$ of the 2D hydrogen atom in
  a magnetic field, for the case $\mathbf{q}=\mathbf{q}_0=\mathbf{0}$
  and the numerical
  coefficients $\mathcal{C}_l^{HI}$ extracted from exact quantum function
  $g_1(\zeta)$ (Eq.~(\protect\ref{scaleddef}))
using harmonic inversion (taking into account multiplicity). The
agreement is excellent for  
the amplitudes and rather nice on the phases, thus emphasizing the
validity of the present theory. That the agreement 
becomes less good for the last orbit only shows the limitations of the
harmonic inversion method, which usually appear on the phase.}
\begin{ruledtabular}
\begin{tabular}{l@{}c@{}c@{}c@{}c}
\multicolumn{1}{c}{Code} & 
\multicolumn{1}{c}{$\mathcal{C}_l$} & 
\multicolumn{1}{c}{$|\mathcal{C}_l^{HI}|$} & 
\multicolumn{1}{c}{Rel. error} &
\multicolumn{1}{c}{$\arg{{C}_l^{HI}}$} \\
\hline
$\overline{13}$     & -0.186\,230\,5 & 0.1864  & $\approx8\times10^{-4}$ 
& 1.002$\times\pi$ \\
$\overline{1243}$    & -0.099\,668\,1 & 0.0995  & $\approx2\times10^{-3}$ 
& 1.01$\times\pi$ \\
$\overline{121343}$   & -0.107\,106\,7 & 0.1072  & $\approx9\times10^{-4}$
& 1.02$\times\pi$ \\
$\overline{12124343}$  & -0.101\,242\,7 & 0.1016  &
$\approx4\times10^{-3}$ & 1.04$\times\pi$  \\ 
$\overline{1212134343}$ & -0.075\,439\,8 & 0.0761  &
$\approx9\times10^{-3}$ & 1.14$\times\pi$
\end{tabular}
\end{ruledtabular}
\end{table}

The Fourier transforms of both functions $g_0(\zeta)$ (upper
plot, solid line) and $g_1(\zeta)$ (lower plot, solid line),
for scaled energy $\epsilon=-0.1$, are displayed in Fig.~\ref{G00TF}.
More precisely, $g_0(\zeta)$ and $g_1(\zeta)$ being known only on a
finite interval $[0,\zeta_{\mathrm{max}}]$, we have plotted the
modulus of their
windowed Fourier transforms, defined as follows:
\begin{equation}
\label{WFTGQQ0}
\begin{split}
F_0(s)&=\frac{6}{(\zeta_{\mathrm{max}})^3}
\int_0^{\zeta_{\mathrm{max}}}d\zeta\,\,\zeta
(\zeta_{\mathrm{max}}-\zeta)g_0(\zeta)e^{-i2\pi s\zeta}, \\
F_1(s)&=\frac{6}{(\zeta_{\mathrm{max}})^3}
\int_0^{\zeta_{\mathrm{max}}}d\zeta\,\,\zeta
(\zeta_{\mathrm{max}}-\zeta)g_1(\zeta)e^{-i2\pi s\zeta}.
\end{split}
\end{equation}
As expected, they depict peaks at the classical actions of
closed orbits, whose trajectories in $(u,v)$ plane have been inserted
in the figure, the black circle corresponding to the nucleus position.
In the figure, the dotted lines corresponds to
the semiclassical estimations of the same functions using the
classical properties given by Table~\ref{tabclosed}. The closed orbits being
either half of a periodic orbit or a periodic orbit, we label a given
close orbit
with the four-disk code of the corresponding periodic
orbit~\cite{EW90,H95}. 

For the leading order in $\hbar$ (upper plot), as expected, the
agreement between 
the quantum results and the semiclassical estimation is
excellent. For the first order $\hbar$ correction, the 
agreement is very good, but one can notice that there is a discrepancy
for the amplitude of the last two peaks. This is not due to errors or
inacurrate calculations in the semiclassical estimation, but rather a
manifestation of the limitations of the Fourier transform. To
emphasize this point, we have used the harmonic inversion to extract,
for each of these orbits, the $\hbar$ correction coefficients 
$\mathcal{C}_l^{HI}$,
from the quantum function $g_1(\zeta)$. The results are compared to
the classical 
calculation $\mathcal{C}_l$ in Table~\ref{tabclosnum}. The agreement
is excellent, the relative error on the amplitude being lower
than $10^{-2}$. As usual, the phase extracted using harmonic
inversion, being the most sensitive quantity, the agreement on the sign
of the $\mathcal{C}_l$, rather nice for the first four orbits, decreases
rapidly. 
Finally, one must mention that this good agreement between quantum and
semiclassical calculations has also been found when considering
quantum Green's functions $G(\mathbf{q},\mathbf{q}_0,2)$ with other
initial or final points.

\subsection{$\hbar$ corrections for
  $\mathrm{Tr}\,\,G(\mathbf{q},\mathbf{q},2)$} 

\begin{figure}
\includegraphics[width=7cm,angle=-90]{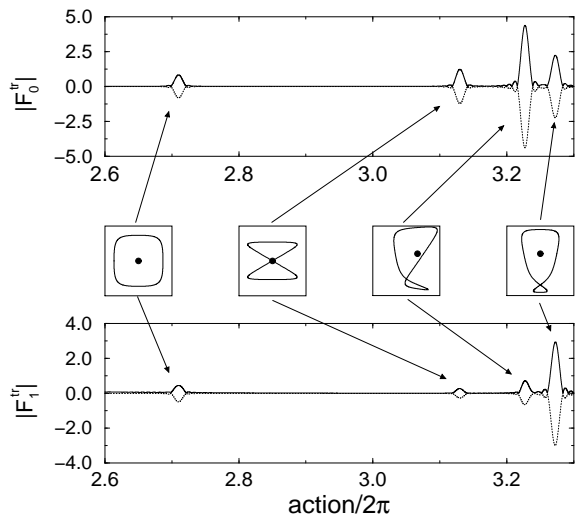}
\caption{\label{TRGTF} Modulus of the windowed Fourier transforms
  $F_0^{\mathrm{tr}}$ (solid line, upper plot) and $F_1^{\mathrm{tr}}$
  (solid line, lower plot), see Eq.~(\protect\ref{WFTTRG}), 
  of the quantum functions $g_0^{\mathrm{tr}}$ (leading order in
  $\hbar$) and $g_1^{\mathrm{tr}}$ (first order $\hbar$ correction), see
  Eq.~(\protect\ref{scaleddef}), associated with the trace of the
  quantum Green's function $\mathrm{Tr}\,\,G(\mathbf{q},\mathbf{q},2)$
in the case of the 2D hydrogen atom in a magnetic field
 (see Sec.~\protect\ref{bidihy} for all details). As expected from
semiclassical 
formula~(\protect\ref{trgreen}), peaks are appearing at action (i.e. 
$\oint\mathbf{p}\,d\mathbf{q}/2\pi$) corresponding to classical periodic
orbits, whose
trajectories in the $(u,v)$ plane are plotted ( the
nucleus being depicted by the black circle). The agreement with the
semiclassical estimation (dotted lines) is excellent, as it is
emphasized by the quantitative 
comparison (using harmonic inversion) displayed by
Table~\protect~\ref{tabpernum}.}
\end{figure}

Still working at scaled energy $\epsilon=-0.1$, Fig.~\ref{TRGTF}
depicts the modulus of the windowed Fourier transforms of
$g_0^{\mathrm{tr}}(\zeta)$ and $g_1^{\mathrm{tr}}(\zeta)$,
$F_0^{\mathrm{tr}}$ (upper plot, solid line), and  $F_1^{\mathrm{tr}}$
(lower plot, solid line),
defined, as previously, as follows:
\begin{equation}
\label{WFTTRG}
\begin{split}
F_0^{\mathrm{tr}}(s)&=\frac{6}{(\zeta_{\mathrm{max}})^3}
\int_0^{\zeta_{\mathrm{max}}}d\zeta\,\,\zeta
(\zeta_{\mathrm{max}}-\zeta)g_0^{\mathrm{tr}}(\zeta)e^{-i2\pi s\zeta}, \\
F_1^{\mathrm{tr}}(s)&=\frac{6}{(\zeta_{\mathrm{max}})^3}
\int_0^{\zeta_{\mathrm{max}}}d\zeta\,\,\zeta
(\zeta_{\mathrm{max}}-\zeta)g_1^{\mathrm{tr}}(\zeta)e^{-i2\pi s\zeta}.
\end{split}
\end{equation}
The trajectories in the $(u,v)$
plane associated with the peaks are also plotted in the
figure. The classical properties of the corresponding periodic orbits
are displayed by Table~\ref{tabperiodic}.  Again the agreement is
excellent between the quantum results (solid lines) and the
semiclassical estimation (dotted lines). The quantitative comparison between
the classical coefficients $\mathcal{C}_l^{\mathrm{tr}}$ and the
values $\mathcal{C}_l^{HI}$
extracted from the quantum function $g^{\mathrm{tr}}(\zeta)$ is 
given in Table~\ref{tabpernum}. The agreement is excellent for
the amplitude of the coefficients  and is
rather good for their phases, which emphasized the validity of the
semiclassical formula developed in the preceding sections, especially
the additional term arising from the Jacobian describing the change
from the Cartesian to local (along the periodic orbit) coordinates
(see Eq.~(\ref{detcarloc})) and
which contributes to a large part of the $\hbar$ correction for the
present orbits.

\begin{table}
\caption{\label{tabperiodic}Classical properties of periodic orbits involved in
  the semiclassical expansion of the trace of the quantum Green's 
  function $\mathrm{Tr}\,\,G(\mathbf{q},\mathbf{q},2)$ of the 
  2D hydrogen atom in a magnetic field. Their trajectories in the
  $(u,v)$ plane are shown in Fig.~\ref{TRGTF}. $S_l^{\mathrm{tr}}$ is
  the 
  reduced action (i.e,. $\oint\mathbf{p}\,d\mathbf{q}/2\pi$), $T_l$ is
  the period, $\mathcal{A}_l^{\mathrm{tr}}$ is the leading semiclassical
amplitude, $\mu_l$ is the Maslov index,
$\mathcal{C}_l^{\mathrm{tr}}$ is the first order $\hbar$ correction,
given by the sum $C_1(T_l)+C_1^{T\rightarrow E}(T_l)$, see Eq.~(\ref{CPTRG}).}
\begin{ruledtabular}
\begin{tabular}{l@{}l@{}l@{}l@{}c}
\multicolumn{1}{c}{Code} & 
\multicolumn{1}{c}{$S_l^{\mathrm{tr}}$} & 
\multicolumn{1}{c}{$T_l^{\mathrm{tr}}$} & 
\multicolumn{1}{c}{$\mathcal{A}_l^{\mathrm{tr}}$} & 
\multicolumn{1}{c}{$\mu_l$}  \\
\hline
$\overline{1234}$  & 2.709\,851\,3 & 6.204\,155\,6 & 0.827\,881\,4 & 4 \\
$\overline{1243}$  & 3.129\,996\,4 & 7.200\,274\,7 & 0.616\,496\,8 & 4  \\
$\overline{12434}$ & 3.227\,168\,1 & 7.541\,640\,6 & 0.548\,479\,1 & 5 \\
$\overline{123434}$& 3.272\,238\,1 & 7.748\,406\,8 & 0.555\,880\,6 & 6  \\
\\
\multicolumn{1}{c}{Code} &
\multicolumn{1}{c}{$C_1(T_l)$} &
\multicolumn{1}{c}{$C_1^{T\rightarrow E}(T_l)$} & 
\multicolumn{1}{c}{$\mathcal{C}_l^{\mathrm{tr}}$} & \\
\hline
$\overline{1234}$   & $-$0.622577 & 0.026912 & $-$0.595665 & \\
$\overline{1243}$   & $\phantom{-}$0.166821 & 0.051665 & 
$\phantom{-}$0.218486 & \\
$\overline{12434}$  & $-$0.203536 & 0.058541 & $-$0.144995 & \\
$\overline{123434}$ & $-$1.41705  & 0.07241  & $-$1.34464  & \\
\end{tabular}
\end{ruledtabular}
\end{table}

\begin{table}
\caption{\label{tabpernum}Numerical comparison between the
  theoretical $\hbar$ corrections $\mathcal{C}_l^{\mathrm{tr}}$ for
  the trace of the quantum
  Green's function $\mathrm{Tr}\,\,G(\mathbf{q},\mathbf{q},2)$ of the 2D
  hydrogen atom in a magnetic field and the numerical
  coefficients $\mathcal{C}_l^{HI}$ extracted from exact quantum function
  $g_1^{\mathrm{tr}}(\zeta)$ (Eq.~(\ref{scaleddef}))
using harmonic inversion (taking into account multiplicity). The
agreement is excellent for  
the amplitudes and rather nice on the phases, thus emphasizing the
validity of the present theory, especially the additional term due to the
transformation from the Cartesian coordinates to the local frame
along the periodic orbit (see Eq.~(\ref{detcarloc})).}
\begin{ruledtabular}
\begin{tabular}{l@{}l@{}l@{}l@{}l}
\multicolumn{1}{c}{Code} & 
\multicolumn{1}{c}{$\mathcal{C}_l^{\mathrm{tr}}$} & 
\multicolumn{1}{c}{$|\mathcal{C}_l^{HI}|$} & 
\multicolumn{1}{c}{Rel. error} &
\multicolumn{1}{c}{$\arg{\mathcal{C}_l^{HI}}$} \\
\hline
$\overline{1234}$    & $-$0.595\,665 & 0.5958  & $\approx2\times10^{-4}$ 
& 1.005$\times\pi$ \\
$\overline{1243}$    & $\phantom{-}$0.218\,486  & 0.2178  & $\approx3\times10^{-3}$ 
& 0.04$\times\pi$ \\
$\overline{12434}$   & $-$0.144\,995 & 0.147   & $\approx1\times10^{-2}$ 
& 0.93$\times\pi$ \\
$\overline{123434}$  & $-$1.344\,64  & 1.347   & $\approx2\times10^{-3}$ 
& 0.98$\times\pi$    
\end{tabular}
\end{ruledtabular}
\end{table}

\section{Conclusion}

In summary, we have explained in this paper how to effectively compute
$\hbar$ corrections in the semiclassical expansions of the propagator
$K(\mathbf{q},\mathbf{q}_0,T)$, its trace $K(T)$, the quantum Green's
function $G(\mathbf{q},\mathbf{q}_0,E)$ and its trace $G(E)$ for chaotic
systems with smooth potential. The method
is based on the classical Green's functions associated to the
relevant trajectories, that is either going from $\mathbf{q}$ to
$\mathbf{q}_0$ in the propagator case or periodic orbits for $K(T)$,
together with adapted boundary conditions. We have shown how all
quantities can be obtained by integrating, 
using the standard Runge-Kutta method, sets of differential equations. We
have also shown that in the derivation of the semiclassical expansion
for $K(T)$ (and thus $G(E)$), starting from the Feynman path integral,
one must take into account additional terms, which affect only $\hbar$
correction coefficients. This is emphasized by the
excellent agreement 
observed when comparing, in the case of the 2D hydrogen atom in
a magnetic field, our theoretical results with the numerical
coefficients extracted from exact quantum data, using the harmonic
inversion. Obviously, there are still many points to be
developed. Besides the few cases, such as self-retracing orbits or
continuous families of orbits, needing specific extensions,
it would be very interesting to understand
how to include continuous and discrete symmetries. Also, going into the
extended phase space $(\mathbf{q},t,\mathbf{p},-E)$~\cite{L31}, it would be
possible to get a better understanding of similarities observed
between the differential sets leading, on one side to the $\hbar$
corrections for the propagator and its trace and, on the other side to
the additional terms arising in the $\hbar$ corrections
for the quantum Green's function and its trace.  

\begin{acknowledgments}
The author thanks D.~Delande for fruitful discussions and
for his kind support during this work, especially for numerous
suggestions which led to the present form.
The author also thanks M.~Ku\'s for useful discussions.
Laboratoire Kastler Brossel is laboratoire de l'Universit{\'e} Pierre et Marie
Curie et de l'Ecole Normale Sup{\'e}rieure, unit{\'e} mixte de
recherche 8552 du CNRS.
\end{acknowledgments}

\appendix*

\section{Few properties of $M(T)$}
\label{appendixb}

In this appendix, we consider an isolated unstable periodic orbit of
period $T$. We shall use the notations $\mathbf{e}_{\parallel}$ and 
$\mathbf{e}_{\perp}$ for the units vectors which are, respectively,
parallel to the flow and perpendicular to the energy shell at the 
initial point.
From Hamilton's equations, we have that 
$M(T)\cdot\mathbf{e}_{\parallel}=\mathbf{e}_{\parallel}$, i.e., 
$\mathbf{e}_{\parallel}$ is an eigenvector of the matrix $M(T)$ for
the eigenvalue 1. The symplectic equation fulfilled by $M(T)$, namely,
$M(T)^{\top}\cdot\Sigma\cdot M(T)=\Sigma$, implies that, if
$\mathbf{e}_i$ and $\mathbf{e}_j$ are two eigenvectors for the
eigenvalues $\lambda_i$ and $\lambda_j$, we have the following
properties:
\begin{equation}
\label{EMT}
\left\{\begin{aligned}
M(T)^{\top}\cdot\left(\Sigma\mathbf{e}_i\right)&=\frac{1}{\lambda_i}
\left(\Sigma\mathbf{e}_i\right), \\
(\lambda_i\lambda_j-1)\mathbf{e}_i^{\top}\Sigma\mathbf{e}_j&=0,
\end{aligned}\right.
\end{equation}
showing thus that $1/\lambda_i$ is an eigenvalue of
$M(T)^{\top}$ and, from that, of $M(T)$.  
In addition, $M(T)$ being a real matrix, $\bar{\lambda}_j$ and
$1/\bar{\lambda}_j$ are also eigenvalues of $M(T)$, so that the
nontrivial eigenvalues (i.e., $\ne 1$) either fall in the
$(\lambda,1/\lambda)$ pair or in quadruplet
$(\lambda,1/\lambda,\bar{\lambda},1/\bar{\lambda})$.
 
In the case of
$\mathbf{e}_i=\mathbf{e}_{\parallel}$, the two preceding
equations~(\ref{EMT}) imply that 
$\mathbf{e}_{\perp}$ is an eigenvector of $M(T)^{\top}$ (but not
necessarily of
$M(T)$) for the eigenvalue 1 and that for every $\lambda_j\neq 1$,
$\mathbf{e}_j$ is orthogonal to $\mathbf{e}_{\perp}$. 
In the basis
$(\mathbf{e}_{\parallel},\mathbf{e}_{\perp},\mathbf{e}_1,\cdots,
\mathbf{e}_{2f-2})$, $M(T)$ entries then read
\begin{equation}
M(T)=\left[\begin{array}{cccccc}
1 & \alpha_{\parallel} & 0 & 0 & \cdots & 0 \\
0 & 1 & 0 & 0 & \cdots & 0 \\
0 & \alpha_1 & \lambda_1 & 0 & \cdots & 0 \\
0 & \alpha_2 & 0 & \lambda_2 & \cdots & 0 \\
\vdots & \vdots & \vdots & \vdots & \ddots & \vdots \\
0 & \alpha_{2f-2} & 0 & 0 & \cdots & \lambda_{2f-2}
\end{array}\right],
\end{equation}
where we have supposed that all eigenvalues are simple. For degenerated
eigenvalues, $M(T)$ would be block diagonal. For a generic periodic
orbit, $\alpha_{\parallel}$ and $\alpha_i$ are nonvanishing
emphasizing thus that $\mathbf{e}_{\perp}$ is not an eigenvector of
$M(T)$. Introducing the vector $\tilde{\mathbf{e}}_{\perp}$ defined as
follows:
\begin{equation}
\tilde{\mathbf{e}}_{\perp}=\mathbf{e}_{\perp}
+\sum_{j=1}^{2f-2}\beta_j\mathbf{e}_j\qquad\text{with}\qquad
\beta_j=\frac{\alpha_j}{1-\lambda_j}
\end{equation}
one immediately gets that
\begin{equation}
M(T)\tilde{\mathbf{e}}_{\perp}=\tilde{\mathbf{e}}_{\perp}+
\alpha_{\parallel}\mathbf{e}_{\parallel}.
\end{equation}
In the case $\alpha_{\parallel}=0$, we have thus found
another eigenvector for the eigenvalue 1, which means that a small
displacement of initial conditions in the $\tilde{\mathbf{e}}_{\perp}$
direction leads to another periodic motion with the same period $T$,
and thus that the periodic orbit is
actually embedded in a continuous family. Indeed, using notations from
Sec.~\ref{Sec:TG}, one can show that
\begin{equation}
\mathbf{X}^{(1)}(0)=-\frac{\|\dot{\mathbf{X}}\|}{\alpha_{\parallel}}
\tilde{\mathbf{e}}_{\perp}
\end{equation}
so that we have
\begin{equation}
\alpha_{\parallel}=\|\dot{\mathbf{X}}\|^2\,\partial_ET.
\end{equation}

In Sec.~\ref{Sec:TG}, one needs to compute derivatives with respect to
the period $T$ of $\det{(m(T)-\openone)}$, whose expression in terms of
the nontrivial eigenvalues of the monodromy matrix reads
\begin{equation}
\det{(m(T)-\openone)}=\prod_{j=1}^{2f-2}(\lambda_j-1).
\end{equation}
Introducing $\mathcal{P}_{\parallel}$ and $\mathcal{P}_{\perp}$ the
projectors on the directions $\mathbf{e}_{\parallel}$ and
$\mathbf{e}_{\perp}$, more precisely
\begin{equation}
\mathcal{P}_{\parallel}=\mathbf{e}_{\parallel}\cdot
\mathbf{e}_{\parallel}^{\top}\qquad\text{and}\qquad
\mathcal{P}_{\perp}=\mathbf{e}_{\perp}\cdot
\mathbf{e}_{\perp}^{\top}
\end{equation}
ones defines the matrix $N(T)$ as follows:
\begin{equation}
\label{NT}
N(T)=M(T)-\left(\openone-\mathcal{P}_{\parallel}-\mathcal{P}_{\perp}\right).
\end{equation}

In the basis
$(\mathbf{e}_{\parallel},\mathbf{e}_{\perp},\mathbf{e}_1,\cdots,
\mathbf{e}_{2f-2})$, using orthogonality between $\mathbf{e}_{\perp}$
and $\mathbf{e}_j$, entries of $N(T)$ read
\begin{equation}
N(T)=\left[\begin{array}{cccccc}
1 & \alpha_{\parallel} & \gamma_1 & \gamma_2 & \cdots & \gamma_{2f-2} \\
0 & 1 & 0 & 0 & \cdots & 0 \\
0 & \alpha_1 & \lambda_1-1 & 0 & \cdots & 0 \\
0 & \alpha_2 & 0 & \lambda_2-1 & \cdots & 0 \\
\vdots & \vdots & \vdots & \vdots & \ddots & \vdots \\
0 & \alpha_{2f-2} & 0 & 0 & \cdots & \lambda_{2f-2}-1
\end{array}\right],
\end{equation}
where $\gamma_j=\mathbf{e}_{\parallel}^{\top}\cdot\mathbf{e}_j$, which
actually could be related to the $\alpha_j$, but this is not necessary
in our case. This shows that the determinant of $N(T)$ is exactly
$\prod_{j=1}^{2f-2}(\lambda_j-1)$. The main advantage of the matrix
$N(T)$ is that its expression~(\ref{NT}) does not involve the
eigenvectors or the eigenvalues of $M(T)$, so that its determinant
can be directly computed, without the diagonalization stage required
when getting $\det{(m(T)-\openone)}$ through the eigenvalues
$\lambda_j$. Furthermore, derivatives of $\ln{\det{N(T)}}$ with respect to
the period $T$ are also straightforward to obtain, knowing derivatives
of $M(T)$ and of $\dot{\mathbf{X}}(T)$, whereas derivatives of
$\lambda_j$ would require the knowledge of those of the
eigenvectors $\mathbf{e}_j$.


\end{document}